\documentclass[twocolumn,twocolappendix]{aastex63}

\usepackage{xcolor}

\newcounter{daggerfootnote}

\definecolor{mycol}{rgb}{0,0,1}

\received{July 18, 2022}
\revised{September 16, 2022}
\accepted{October 5, 2022}
\submitjournal{\apjs}

\shorttitle{ALCS: CHArGE \textit{HST} and \textit{Spitzer} photometry}
\shortauthors{Kokorev et al.}

\usepackage{longtable}
\usepackage{lipsum}
\usepackage{amsmath}
\usepackage[normalem]{ulem}
\usepackage[para]{threeparttable}

\begin{document}



\title{ALMA Lensing Cluster Survey: $HST$ and $Spitzer$ Photometry of 33 Lensed Fields Built with CHArGE}


\correspondingauthor{Vasily Kokorev}
\email{vasilii.kokorev@nbi.ku.dk}

\author[0000-0002-5588-9156]{V. Kokorev}
\affiliation{Cosmic Dawn Center (DAWN), Jagtvej 128, DK2200 Copenhagen N, Denmark}
\affiliation{Niels Bohr Institute, University of Copenhagen, Blegdamsvej 17, DK2100 Copenhagen \O, Denmark}

\author[0000-0003-2680-005X]{G. Brammer}
\affiliation{Cosmic Dawn Center (DAWN), Jagtvej 128, DK2200 Copenhagen N, Denmark}
\affiliation{Niels Bohr Institute, University of Copenhagen, Blegdamsvej 17, DK2100 Copenhagen \O, Denmark}

\author[0000-0001-7201-5066]{S. Fujimoto}
\affiliation{Cosmic Dawn Center (DAWN), Jagtvej 128, DK2200 Copenhagen N, Denmark}
\affiliation{Niels Bohr Institute, University of Copenhagen, Blegdamsvej 17, DK2100 Copenhagen \O, Denmark}

\author[0000-0002-4052-2394 ]{K. Kohno}
\affiliation{Institute of Astronomy, Graduate School of Science, The University of Tokyo, 2-21-1 Osawa, Mitaka, Tokyo 181-0015, Japan}
\affiliation{Research Center for the Early Universe, School of Science, The University of Tokyo, 7-3-1 Hongo, Bunkyo-ku, Tokyo 113-0033, Japan}


\author[0000-0002-4872-2294]{G. E. Magdis}
\affiliation{Cosmic Dawn Center (DAWN), Jagtvej 128, DK2200 Copenhagen N, Denmark}
\affiliation{DTU-Space, Technical University of Denmark, Elektrovej 327, DK2800 Kgs. Lyngby, Denmark}
\affiliation{Niels Bohr Institute, University of Copenhagen, Blegdamsvej 17, DK2100 Copenhagen \O, Denmark}

\author[0000-0001-6477-4011]{F. Valentino}
\affiliation{Cosmic Dawn Center (DAWN), Jagtvej 128, DK2200 Copenhagen N, Denmark}
\affiliation{Niels Bohr Institute, University of Copenhagen, Blegdamsvej 17, DK2100 Copenhagen \O, Denmark}

\author[0000-0003-3631-7176]{S. Toft}
\affiliation{Cosmic Dawn Center (DAWN), Jagtvej 128, DK2200 Copenhagen N, Denmark}
\affiliation{DTU-Space, Technical University of Denmark, Elektrovej 327, DK2800 Kgs. Lyngby, Denmark}
\affiliation{Niels Bohr Institute, University of Copenhagen, Blegdamsvej 17, DK2100 Copenhagen \O, Denmark}

\author[0000-0001-5851-6649]{P. Oesch}
\affiliation{Department of Astronomy, University of Geneva, Chemin Pegasi 51, 1290 Versoix, Switzerland}
\affiliation{Cosmic Dawn Center (DAWN), Jagtvej 128, DK2200 Copenhagen N, Denmark}

\author[0000-0002-2951-7519]{I. Davidzon}
\affiliation{Cosmic Dawn Center (DAWN), Jagtvej 128, DK2200 Copenhagen N, Denmark}
\affiliation{Niels Bohr Institute, University of Copenhagen, Blegdamsvej 17, DK2100 Copenhagen \O, Denmark}

\author[0000-0002-8686-8737]{F. E. Bauer}
\affil{Instituto de Astrof{\'{\i}}sica, Facultad de F{\'{i}}sica, Pontificia Universidad Cat{\'{o}}lica de Chile, Campus San Joaquín, Av. Vicuña Mackenna 4860, Macul Santiago, Chile, 7820436} 
\affil{Centro de Astroingenier{\'{\i}}a, Facultad de F{\'{i}}sica, Pontificia Universidad Cat{\'{o}}lica de Chile, Campus San Joaquín, Av. Vicuña Mackenna 4860, Macul Santiago, Chile, 7820436} 
\affil{Millennium Institute of Astrophysics, Nuncio Monse{\~{n}}or S{\'{o}}tero Sanz 100, Of 104, Providencia, Santiago, Chile} 

\author[0000-0001-7410-7669]{D. Coe}
\affiliation{Space Telescope Science Institute, 3700 San Martin Drive, Baltimore, MD 21218, USA}

\author{E. Egami}
\affiliation{Steward Observatory, University of Arizona, 933 N. Cherry Avenue, Tucson, 85721, USA}

\author[0000-0003-3484-399X]{M. Oguri}
\affiliation{Research Center for the Early Universe, Graduate School of Science, The University of Tokyo, 7-3-1 Hongo, Bunkyo-ku, Tokyo 113-0033, Japan}
\affiliation{Center for Frontier Science, Chiba University, 1-33 Yayoi-cho, Inage-ku, Chiba 263-8522, Japan}
\affiliation{Kavli Institute for the Physics and Mathematics of the Universe (WPI), The University of Tokyo, 5-1-5 Kashiwanoha, Kashiwa-shi, Chiba, 277-8583, Japan}

\author[0000-0002-1049-6658]{M. Ouchi}
\affiliation{National Astronomical Observatory of Japan, 2-21-1 Osawa, Mitaka, Tokyo 181-8588, Japan}
\affiliation{Institute for Cosmic Ray Research, The University of Tokyo, 5-1-5 Kashiwanoha, Kashiwa, Chiba 277-8582, Japan}
\affiliation{Kavli Institute for the Physics and Mathematics of the Universe (WPI), University of Tokyo, Kashiwa, Chiba 277-8583, Japan}

\author[0000-0002-9365-7989]{M. Postman}
\affiliation{Space Telescope Science Institute, 3700 San Martin Dr., Baltimore,
MD 21218, USA}

\author[0000-0001-5492-1049]{J. Richard}
\affiliation{Univ Lyon, Univ Lyon1, Ens de Lyon, CNRS, Centre de Recherche Astrophysique de Lyon UMR5574, F-69230 Saint-Genis-Laval, France}

\author[0000-0002-3405-5646]{J.-B. Jolly}
\affiliation{Department of Space, Earth and Environment, Chalmers University of Technology, Onsala Space Observatory, 439 92 Onsala,
Sweden}
\affiliation{Max-Planck-Institut für extraterrestrische Physik, 85748 Garching, Germany}

\author[0000-0002-7821-8873]{K. K. Knudsen}
\affiliation{Department of Space, Earth and Environment, Chalmers University of Technology, Onsala Space Observatory, 439 92 Onsala,
Sweden}

\author[0000-0002-4622-6617]{F. Sun}
\affiliation{Steward Observatory, University of Arizona, 933 N. Cherry Avenue, Tucson, 85721, USA}

\author[0000-0003-1614-196X]{J. R. Weaver}
\affil{Department of Astronomy, University of Massachusetts, Amherst, MA 01003, USA}
\affiliation{Cosmic Dawn Center (DAWN), Jagtvej 128, DK2200 Copenhagen N, Denmark}


\author{Y. Ao}
\affiliation{Purple Mountain Observatory and Key Laboratory for Radio Astronomy, Chinese Academy of Sciences, Nanjing, China}
\affiliation{School of Astronomy and Space Science, University of Science and Technology of China, Hefei, China}

\author[0000-0002-7892-396X]{A. J. Baker}
\affiliation{Department of Physics and Astronomy, Rutgers, the State University of New Jersey, 136 Frelinghuysen Road, Piscataway, NJ 08854-8019, USA}
\affiliation{Department of Physics and Astronomy, University of the Western Cape, Robert Sobukwe Road, Bellville 7535, South Africa}

\author[0000-0002-7908-9284]{L. Bradley}
\affiliation{Space Telescope Science Institute, 3700 San Martin Drive, Baltimore, MD 21218, USA}

\author[0000-0001-8183-1460]{K. I. Caputi}
\affiliation{Kapteyn Astronomical Institute, University of Groningen, P.O. Box 800, 9700AV Groningen, The Netherlands}
\affiliation{Cosmic Dawn Center (DAWN), Jagtvej 128, DK2200 Copenhagen N, Denmark}

\author[0000-0003-0348-2917]{M. Dessauges-Zavadsky}
\affiliation{Observatoire de Gen\'{e}ve, Universit\'{e} de Gen\'{e}ve, Versoix, Switzerland}

\author[0000-0002-8726-7685]{D. Espada}
\affiliation{Departamento de F\'{i}sica Te\'{o}rica y del Cosmos, Campus de Fuentenueva, Edificio Mecenas, Universidad de Granada, E-18071, Granada, Spain}
\affiliation{Instituto Carlos I de F\'{i}sica Te\'{o}rica y Computacional, Facultad de Ciencias, E-18071, Granada, Spain}

\author[0000-0001-6469-8725]{B. Hatsukade}
\affiliation{Institute of Astronomy, Graduate School of Science, The University of Tokyo, 2-21-1 Osawa, Mitaka, Tokyo 181-0015, Japan}

\author[0000-0002-6610-2048]{A. M. Koekemoer}
\affiliation{Space Telescope Science Institute, 3700 San Martin Dr., Baltimore, MD 21218, USA}

\author[0000-0002-8722-516X]{A. M. Mu\~noz Arancibia}
\affiliation{Millennium Institute of Astrophysics (MAS), Nuncio Monse\~nor S\'otero Sanz 100, Providencia, Santiago, Chile}
\affiliation{Center for Mathematical Modeling, Universidad de Chile, Beauchef 851, Santiago 8320000, Chile}

\author[0000-0002-2597-2231]{K. Shimasaku}
\affiliation{Department of Astronomy, Graduate School of Science, The University of Tokyo, 7-3-1 Hongo, Bunkyo-ku, Tokyo 113-0033, Japan}
\affiliation{Research Center for the Early Universe, Graduate School of Science, The University of Tokyo, 7-3-1 Hongo, Bunkyo-ku, Tokyo
113-0033, Japan}

\author[0000-0003-1937-0573]{H. Umehata}
\affiliation{Institute for Advanced Research, Nagoya University, Furocho, Chikusa, Nagoya 464-8602, Japan}
\affiliation{Department of Physics, Graduate School of Science, Nagoya University, Furocho, Chikusa, Nagoya 464-8602, Japan}

\author{T. Wang}
\affiliation{Key Laboratory of Modern Astronomy and Astrophysics (Nanjing University), Ministry of Education, Nanjing 210093, China}

\author[0000-0003-2588-1265]{W.-H. Wang}
\affiliation{Institute of Astronomy and Astrophysics, Academia Sinica, No 1, Sec 4, Roosevelt Rd., Taipei City, 10617, Taiwan}

\begin{abstract}
We present a set of multi-wavelength mosaics and photometric catalogs in the ALMA lensing cluster survey (ALCS) fields. The catalogs were built by reprocessing of archival data from the CHArGE compilation, taken by the \textit{Hubble Space Telescope} (\textit{HST}) in the RELICS, CLASH and Hubble Frontier Fields. Additionally we have reconstructed the \textit{Spitzer} IRAC 3.6 and 4.5\,$\mu$m mosaics, by utilising all the available archival IRSA/SHA exposures. 
To alleviate the effect of blending in such a crowded region, we have modelled the \textit{Spitzer} photometry by convolving the \textit{HST} detection image with the \textit{Spitzer} PSF
using the novel \textsc{golfir} software.
The final catalogs contain 218,000 sources, covering a combined area of 690\,arcmin$^2$, a factor of $\sim \times 2$ improvement over the currently existing photometry. A large number of detected sources is a result of re-processing of all available and sometimes deeper exposures, in conjunction with a combined optical-NIR detection strategy. These data will serve as an important tool in aiding the search of the sub-mm galaxies in future ALMA surveys, as well as follow ups of the \textit{HST} dark and high-$z$ sources with \textit{JWST}.
Coupled with the available \textit{HST} photometry the addition of the 3.6 and 4.5\,$\mu$m bands will allow us to place a better constraint on photometric redshifts and stellar masses of these objects, thus giving us an opportunity to identify high-redshift candidates for spectroscopic follow ups and answer the important questions regarding the epoch of reionization and formation of first galaxies. The mosaics, photometric catalogs and the best-fit physical properties are publicly available. \protect\footnote{\url{https://github.com/dawn-cph/alcs-clusters}}

\end{abstract}

\keywords{catalogs –- galaxies: evolution –- galaxies: high-redshift –- galaxies: ISM –-  submillimeter: ISM: photometry -- methods: observational –- techniques: photometric}

\section{Introduction} \label{sec:intro}
The emergence of large multi-wavelength photometric surveys has allowed us to conduct detailed studies of galaxy formation and evolution across cosmic time by observing statistically significant population of galaxies. In particular, the investment of thousands of orbits of \textit{Hubble Space Telescope} (\textit{HST}) and \textit{Spitzer Space Telescope} (\textit{Spitzer}) time has cemented their unprecedented imaging legacy and enabled us to revolutionize our understanding of both observational cosmology and galaxy evolution.
For example, these unique capabilities allowed us to capture the accelerating expansion of the Universe \citep{riess04}, and have helped demonstrate that the majority of star formation took place within a relatively short time span, in the epoch at $1<z<3$ (see e.g. \citealt{hopkinsbeacom06,bouwens07}). 
\par
More recently, the advantages of space-based observations have become particularly pronounced in the search for high redshift galaxies, with the combined efforts of the very sensitive Wide Field Camera 3 (WFC3) - onboard \textit{HST} and the ultra deep \textit{Spitzer} Infrared Array Camera (IRAC) imaging. The remarkable wavelength coverage of these instruments has helped us push the observational frontier to the end of the cosmic epoch of reionization at $z\sim 7-8$, some 700 Myr from the Big Bang, and toward the epoch beyond $z\sim10$, where the formation of the first galaxies has taken place. A number of large deep extragalactic blank field surveys has now led to the discovery of a significant and statistically meaningful number of galaxies at $z\sim 7-8$ (e.g. \citealt{mclure13,finkelstein15,bouwens15}), an ever growing sample of $z\sim9-11$ candidates \citep{ellis13,oesch13,oesch14,bouwens16,calvi16}, and even the most distant galaxy discovered to date at $z=11.1$ \citep{oesch16,jiang21}. 
The most staggering and impactful discoveries of high-$z$ galaxies have however been made within lensing cluster fields, which include Hubble Frontier Fields (HFF;  \citealt{lotz17}), the Reionization Lensing Cluster Survey (RELICS; \citealt{coe19}), and the Cluster Lensing And Supernova survey with Hubble (CLASH; \citealt{postman12}). All three combine the power of \textit{HST} and \textit{Spitzer}  observations and a strong gravitational lensing potential of massive galaxy clusters, to produce the deepest available observations of high-$z$ galaxies lensed by clusters ever obtained (e.g. see \citealt{zheng12,coe13,bradley14,schmidt14,zitrin14,infante15,ishigaki15,kawamata15,oesch15,mcleod15,hashimoto18,hoag18,strait20}).
\par
The redshift estimates of these objects still largely rely on spectral energy distribution (SED) fitting of broad-band photometry, and spectroscopically confirmed samples remain limited. The SED fitting photometric redshift technique is largely leveraged on the correct identifications of either the Lyman or Balmer breaks, at 912 \AA\, and 3640 \AA\, respectively, in the stellar continuum. At $z\sim9-10$ the \textit{Spitzer}/IRAC targets the $\sim 3000-4000$\,\AA\, rest frame continuum and as such can greatly aid in removing the low-redshift interlopers from the high-$z$ samples. Moreover, even for spectroscopically confirmed objects the \textit{Spitzer} observations are essential for conducting robust measurements of the stellar population parameters, such as stellar mass ($M_*$), dust attenuated star formation rate (SFR), and extinction ($A_V$) \citep{gonzalez11,ryan14,salmon15}. The existing data has already lead to implications that the first main episodes of star-formation have taken place $\sim 250$ Myr after the Big Bang \citep{hashimoto18}. 
\par
Some questions regarding star formation, are however yet to be answered. The rest-frame UV and FIR observations conducted during the last decade \citep{lefloch05,lefloc2009},
present us with a view of the Universe where the star formation has already reached its peak at $z\sim 2-3$ and is now declining (e.g. \citealt{madau14,zavala21}). Studies of the SFR density (SFRD) also show a growing disparity between the contribution of dust obscured SFR, measured from the IR data, and the unobscured SFR, measured from the UV - optical data.
This, in return, might imply that the early Universe was less dusty, aligning with some of our predictions regarding the time-scale and mechanisms of dust production. On the other hand, blind  Atacama Large Millimeter/submillimeter Array (ALMA) studies of galaxies at $z\sim 2 -6 $ \citep{wang16,wang19,yamaguchi19,gruppioni20,umehata20,manning21,sun22} reveal a population of optically dark, dusty sources, which contribute ten times more towards the SFRD, than similarly bright galaxies with a rest-frame UV detection, and reside within centres of the most massive overdensities \citep{zhou21}. The ubiquity of such systems, can potentially create obstacles in our understanding of the true numbers of massive galaxies, SFRD in the early Universe and challenge our current understanding of galaxy formation. As a result, our ability to correctly recover the total SFR comes down to the detection of high-$z$ dusty galaxies, such as ``\textit{HST} - dark'', or optically dark sources.
\par
The complete dust obscuration of UV-optical emission, makes the detection and identification of these objects a significant challenge, even when sufficient MIR and FIR coverage are available. As a result the only reliable way to observe such objects are blind field studies with ALMA or the NOrthern Extended Millimeter Array (NOEMA). ALMA in particular, has been the primary tool driving the discovery of new faint sub-mm galaxies, ($S_{\rm 1.2 mm} \sim 0.02 -1$ mJy), which are substantially fainter compared to the traditionally observed sub-mm galaxies (SMGs) (see e.g. \citealt{hatsukade13,hatsukade16,ono14,carniani15,oteo16,fujimoto16,aravena16,dunlop17,gonzalezlopez17,franco18,sun22}. Over the last few years, ALMA observations of faint 1.2 mm sources have been able to derive ALMA mm counts down to depths of $\sim 0.02$ mJy \citep{carniani15,fujimoto16,hatsukade16,aravena16}, however, despite such depths, the origin of the cosmic infrared background (CIB), and therefore the majority of SFRD still remains hidden. It has quickly become apparent that deeper ALMA observations are absolutely imperative in order to separate and resolve the remained of the CIB into discrete sources, in order to study their individual properties. The most efficient way to complete this puzzle is to observe a sufficiently large number of lensing clusters using ALMA.
\par
The ALMA Lensing Cluster Survey (ALCS; Project ID: 2018.1.00035.L; PI: K. Kohno) aims to do exactly that. At the moment, it is the largest, by area, among other ALMA surveys targeting clusters of galaxies. Combined with previous ALMA observations, the survey covers a total of 33 massive galaxy clusters. The observations aim to provide an in-depth look of the high magnification regions within the cluster fields, and, in particular, target dust-continuum-selected and line-emitting high-$z$ galaxies. The main science goal of the survey is to examine the faint-end slope of the 1.2 mm source counts and to provide the best estimate for the cosmic infrared background (CIB) at that wavelength.
The typical galaxies contributing to the CIB at 1.2 mm are intrinsically faint (see e.g. \citealt{fujimoto16}), however in conjunction with the rich \textit{HST} and \textit{Spitzer}/IRAC data-sets covering the field, the survey aims to reveal the fundamental physical properties of the $S_{\rm 1.2 mm}< 0.1$ mJy galaxies, such as stellar masses and IR based SFRs (see \citealt{sun22}). 
\par
All ALCS clusters have been previously imaged with \textit{HST}, and \textit{Spitzer}/IRAC, enabling accurate positions and other quantities derived from the photometry.  In this work we describe the re-processing of all archival \textit{HST}, and \textit{Spitzer}/IRAC mosaics, covering the ALCS. We perform careful aperture photometry of all \textit{HST} sources, and use them as priors to model and fit the flux densities for the sources in the blended IRAC maps. The final images and catalogs will then act as a powerful tool to establish better constraints on photometric redshifts and physical properties of these objects. In addition, this allows us to identify high-redshift candidates for spectroscopic follow ups and answer the important questions regarding the epoch of reionization and formation of first galaxies.
\par
The provided catalogs include \textit{HST} and \textit{Spitzer} photometry, photometric redshifts, and stellar population properties, for each field included in ALCS (similarly to the ASTRODEEP collaboration \citet{astrodeep1,astrodeep2} and the HFF-DeepSpace catalogs \citet{shipley18}, albeit following a different methodology). Moreover, the public release of our data is complemented by all the new \textit{HST}/ACS, \textit{HST}/WFC3 and IRAC mosaics, including detection images, models, residuals and segmentation maps.
\par
The paper is organised as follows. In \autoref{sec:sample} we list the data sets used in this work, and describe the creation of new \textit{Spitzer}/IRAC mosaics. In \autoref{sec:phot} we describe the high- and low-resolution photometry algorithms. In \autoref{sec:cat} we describe the catalog format, ALMA counterparts, quality flags and present the quality and consistency check for our catalog. In \autoref{sec:param} we describe the spectral energy distribution fitting of our photometry, resultant photometric redshifts, stellar population parameters, and rest-frame colors. Finally our main conclusions and summary are given in \autoref{sec:conc}.
\par
Throughout this paper we assume a flat $\Lambda$CDM cosmology with $\Omega_{\mathrm{m},0}=0.3$, $\Omega_{\mathrm{\Lambda},0}=0.7$ and H$_0=70$ km s$^{-1}$ Mpc$^{-1}$, and a \citet{chabrier} initial mass function (IMF).
All magnitudes in this paper are expressed in the AB system \citep{oke74}, for which a flux $f_\nu$ in Jy ($10^{-23}$~erg~cm$^{-2}$ s$^{-1}$ Hz$^{-1}$) corresponds to AB$=23.9-2.5\,\log_{10}(f_\nu/\mu{\rm Jy})$.

\section{Data Sets} \label{sec:sample}
The ALCS is a Cycle 6 ALMA large program targeting 33 lensing cluster fields in Band 6 ($\bar \lambda=1.15$ mm/$\bar \nu=260$ GHz). In total ALCS covers an area of 134 arcmin$^2$, with synthesized beam response smaller than 0\farcs5, reaching a depth of 70 $\mu$Jy ($1\sigma$). The sample is designed to be contained within the  best-studied massive clusters also imaged in \textit{HST} programs. 
More specifically, ALCS includes 5 clusters from HFF \citep{lotz17}, plus the new BUFFALO observations \citep{steinhardt20}, 16 clusters from the RELICS \citep{coe19}, and 12 clusters from CLASH \citep{postman12}. 
\par
We list a few major motivations in constructing a uniform $HST$/$Spitzer$ mosaics in ALCS fields as follows. The vast majority of objects within ALCS are only continuum detected, ruling out redshift constraints from spectra. However, the rich UV-optical and NIR treasury data already collected within these cluster fields, will allow for the derivation of photometric redshifts for a vast majority of faint sub-mm galaxies. This will allow the derivation of the dust-based FIR luminosity functions, and give upper limits on its evolution at $z>3$, something that has been out of reach for $Spitzer$ and $Herschel$.
\par
The ALCS is also expected to detect the ionized carbon ([CII]) and carbon monoxide (CO) emission lines, facilitating the studies of the ISM for a unique sample of faint galaxies magnified by lensing clusters. Moreover, even for non-detections the mm-properties of various classes of SFGs can be extracted through stacking, facilitated by the presence of both \textit{HST} and IRAC priors within these lensed fields (e.g. see Guerrero et al. 2022 in prep. and Jolly et al. 2022 in prep.).
\par
An additional emphasis of the survey is to detect and characterize the magnified ALMA continuum sources, without \textit{HST} counterparts, i.e. the intrinsically faint, ``\textit{HST} - dark'', ALMA sources. The majority of these objects also have faint counterparts in the $Spitzer$/IRAC bands, with the measured IRAC to 1.2 mm flux density ratios \citep{sun22} pointing towards these sources being either distant $z \gtrsim 4$ galaxies, or massive BCG progenitors at $z\sim4$, similar to the objects presented in \citet{wang16,wang19,yamaguchi19,gruppioni20,umehata20}.

\subsection{CHArGE \textit{HST} Imaging}
The Complete Hubble Archive for Galaxy Evolution (CHArGE) is a novel initiative that performs uniform processing of all archival Hubble Space Telescope imaging and slitless spectroscopy observations relevant for studying distant galaxies (e.g., reasonably high galactic latitudes, avoiding large foreground galaxies, including WFC3/IR).
The data were processed with the \textsc{grizli} pipeline \citep{grizli}, which creates filter mosaics for all ACS, WFC3/UVIS and WFC3/IR exposures that cover a given area of the sky (e.g., an ALCS field). It is worth noting that the produced catalogs span, for each cluster, an area multiple times larger than that of the ALCS ALMA maps. The overlapping exposures are then broken into discrete ``visit'' associations, where the grouping is done for a given filter for data that were taken in a single target acquisition. These associations generally share the same spacecraft orientation and zodiacal sky background.
\par
Within a single visit, all exposures are aligned to each other using high S/N sources detected in them, allowing the relative \texttt{x} and \texttt{y} coordinates between exposures to shift, until the best match is found. These are analogous to \texttt{DrizzlePac TweakShifts} and are generally a fairly small fraction of a pixel for dither offsets within a single orbit and a few tenths of a pixel between subsequent orbits that share the same initial target acquisition.
\par
A source catalog is created from a preliminary mosaic generated from the visit exposures and aligned (shift, rotation, scale) to an astrometric reference catalogue. Generally this is PanSTARRS DR1 \citep{PS1} as it is well aligned to the Gaia DR2 \citep{gaia1,gaia2}, but has a higher source density than the bright Gaia stars alone. 
\par
The final fine alignment is performed simultaneously optimising a) the alignment between all individual visit catalogs, and b) Gaia DR2 stars with proper motions projected to each visit observation epoch. This ensures robust internal alignment of the HST images for matched-aperture photometry and the final absolute astrometric precision is generally $<100$ mas. 
\par
A pedestal sky background of each exposure is estimated in the \texttt{AstroDrizzle} preparation of each visit association. A smooth background is subtracted from each visit mosaic to remove gradients that can then appear as sharp discontinuities in the final combined filter mosaics. This background is estimated with the \textsc{sep} \citep{sep}, a \textsc{Python} implementation of \textsc{SourceExtractor} \citep{sextractor} with \texttt{BACK\_FILTERSIZE = 3} and \texttt{BACK\_SIZE} = 32 arcsec. While the background estimation includes a mask for detected sources, it can include extended structure for very large, bright galaxies and inter-cluster light (ICL) in the ALCS cluster fields.
\par
Final rectified mosaics combining all exposures in each available filter are created with \texttt{AstroDrizzle}. All WFC3/IR mosaics are created with 0\farcs1 pixels, while the ACS and UVIS optical/UV images are drizzled with 0\farcs05 pixels on a grid that subsamples the IR mosaic $2\times2$. Both optical/UV and IR mosaics are drizzled with \texttt{pixfrac = 0.33}. The \texttt{sci} (science) and \texttt{wht} (inverse variance weights) mosaics are provided for each filter.
\par
While this approach might not necessarily result in the best reconstruction of the undersampled \textit{HST} PSFs, but the larger pixels ensure more uniform weights across the diversity of dither coverage across the survey fields (e.g., hundreds of exposures for the Frontier Fields but as few as 2–4 exposures for some pre-RELICS filters). The larger pixels and small \texttt{pixfrac} result in lower correlated noise between adjacent pixels, and therefore the inverse variance maps are a more reliable estimate of the pixel variances for, e.g., aperture photometry.
\par
The units of the filter mosaics are in electrons/s, with the photometric calibration to cgs units provided in the \texttt{PHOTFLAM} ($f_{\lambda}$) and \texttt{PHOTFNU} ($f_{\nu}$) header keywords of each mosaic. We additionally provide PSF models for each IR filter using the effective PSF models described in \citet{anderson_king00}. The sources of all \textit{HST} data used for this work are listed in \autoref{sec:appendix_0}.

\begin{figure*}
\begin{center}
\includegraphics[width=1\textwidth]{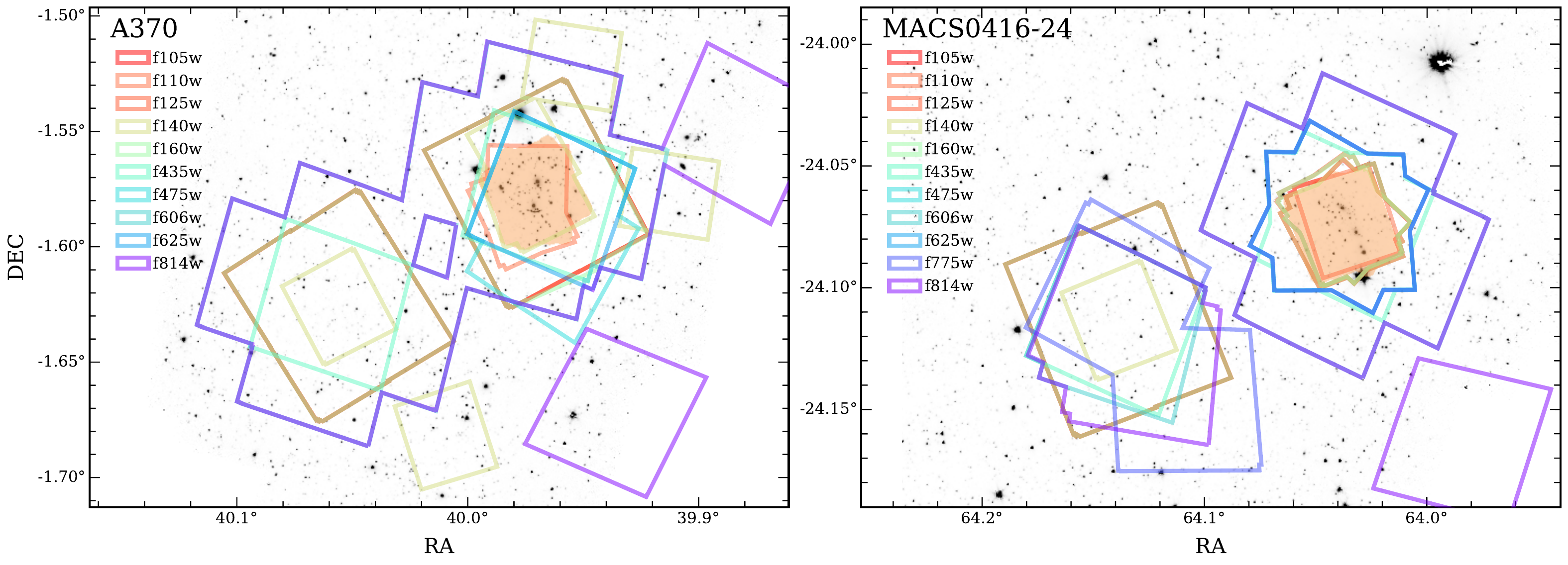}
\includegraphics[width=1\textwidth]{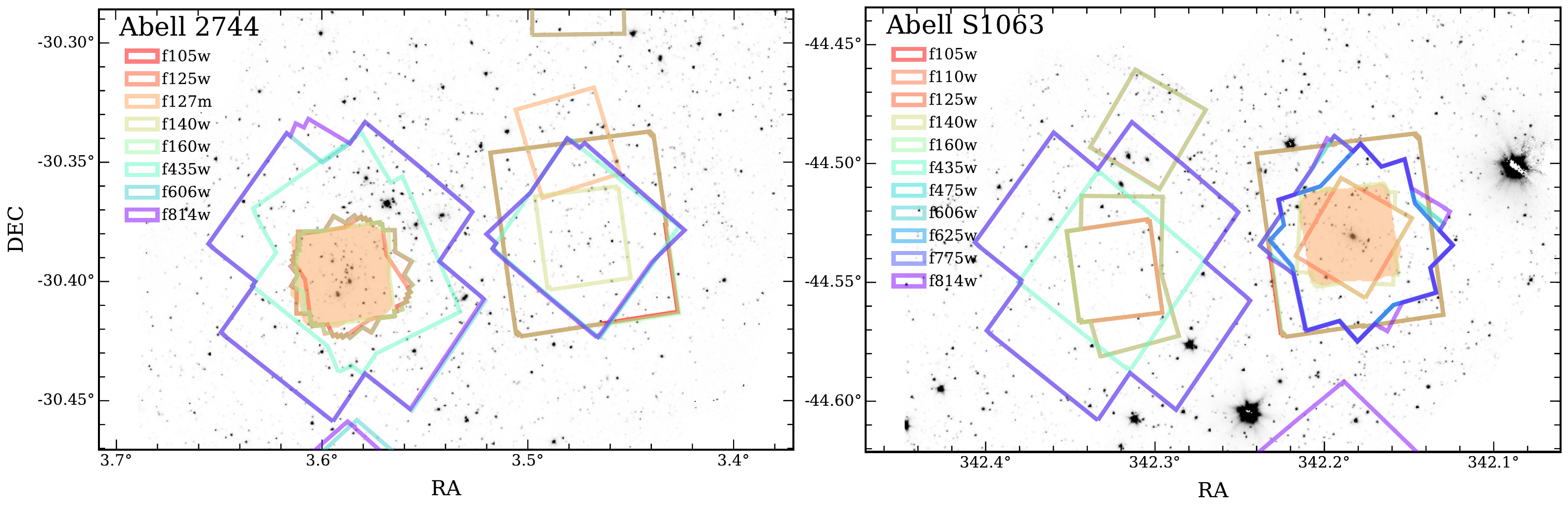}
\caption{Footprint of the \textit{HST} observations, superimposed on the $Spitzer$/IRAC 3.6 $\mu$m data of four Frontier Field Clusters \citep{lotz17}. The ALCS band 6 ALMA coverage is shown as the shaded orange areas. The images shown cover both the cluster and the parallels of the HFF.}
\label{fig:footprint}
\end{center}
\end{figure*}

\subsection{IRAC imaging}

We begin by collecting all the Basic Calibrated Data (BCD) exposures (pBCD), from the \textit{Spitzer} SHA archive for each field in the ALCS. These include IRAC exposures in HFF (Spitzer Frontier Fields; \citealt{spitzer_hff}), RELICS (Spitzer Reionization Lensing Cluster Survey; \citealt{spitzer_relics} ) and CLASH \citep{spitzer_clash}. Our \textit{Spitzer} data only include \textit{Spitzer}/IRAC 3.6 and 4.5 $\mu$m, as our fields of interest do not have a uniform coverage in 5.8 and 8 $\mu$m. 
Similarly to how we process the \textit{HST} CHArGE data, we perform relative alignment of all exposures in the IRAC Astronomical Observation Requests (AOR), and then align AOR mosaics to GAIA DR2. A larger size of IRAC FoV is generally beneficial, as it allows for a sufficient number of reference sources to be used for alignment.\par
The IRAC background is removed by creating a master background image for each AOR with detected sources masked.
Finally, the complete IRAC mosaics are aligned to the \textit{HST} pixel grid and drizzled with 0\farcs5 pixels and the pixel fraction parameter \texttt{pixfrac} set to 0.2. \par
With full knowledge of the individual pBCD exposures that contribute at any location in the final mosaic, we can generate robust models of the IRAC PSF that fully account for the diversity of depth and detector position angle across the mosaic. These position-dependent PSFs are used for the IRAC model-based photometry, and can be re-generated using \textsc{golfir} \citep{golfir}.

\begin{figure}
\begin{center}
\includegraphics[width=.5\textwidth]{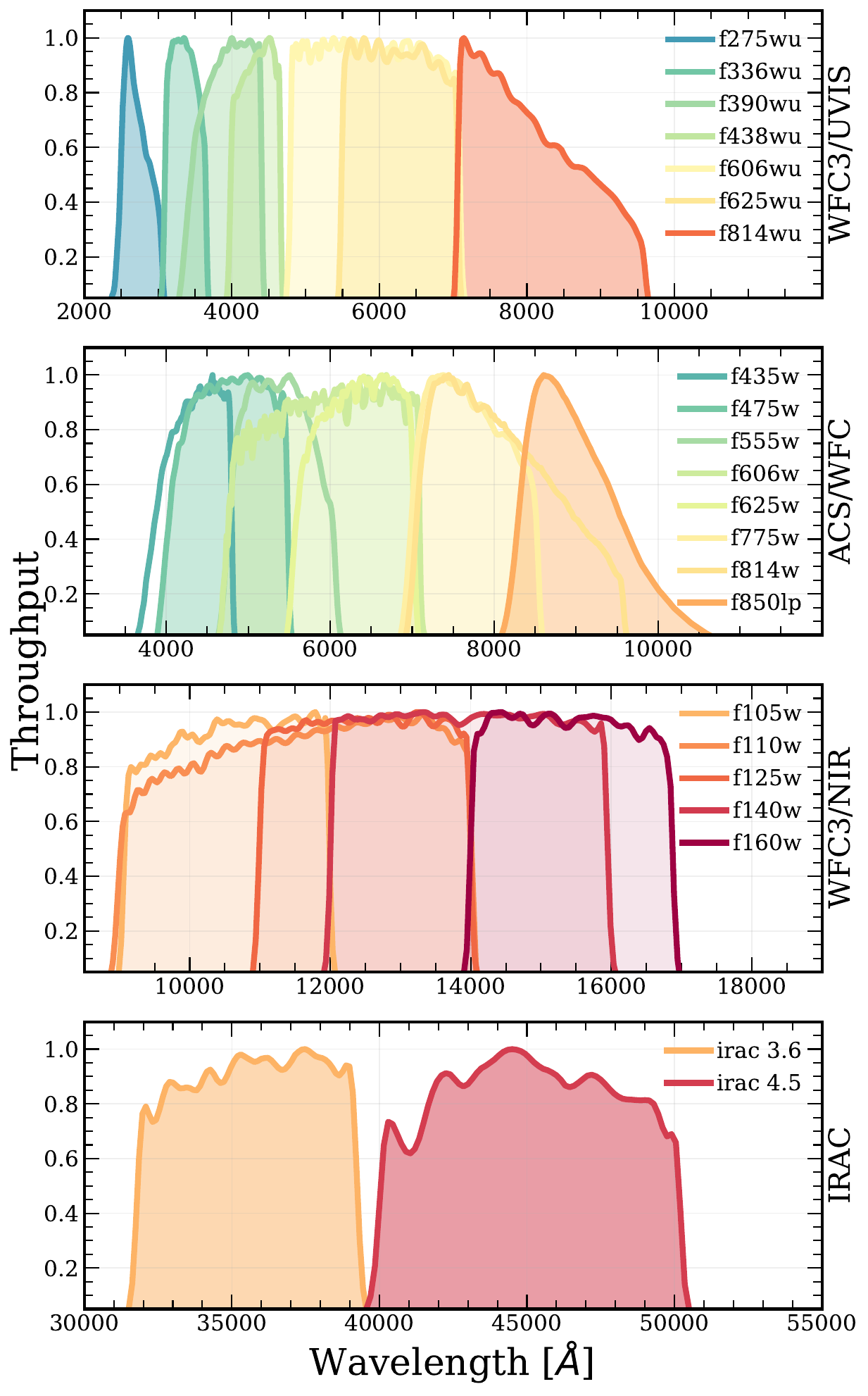}
\caption{The arrangement of all available \textit{HST} filters within ALCS clusters. These filters include (top to bottom): seven broad-bands from WFC3/UVIS (e.g. f275wu), eight bands from ACS/WFC, five broad-bands from WFC3/NIR and two bands from \textit{Spitzer}/IRAC.}
\label{fig:filters}
\end{center}
\end{figure}

\section{Photometry} \label{sec:phot}
During the generation of this catalogue, we have treated the highly resolved ($HST$/ACS and WFC3) data differently from the low-resolution $Spitzer$/IRAC data.
In \autoref{fig:footprint} we show footprints for all the available \textit{HST} and \textit{Spitzer} photometry in four Frontier Fields. We additionally highlight the area covered by ALCS. An array of filter response curves for all available bands in presented in \autoref{fig:filters}.

\subsection{Source detection} \label{s:detection}
The survey fields studied here are crowded with galaxy cluster members and diffuse intra-cluster light that present a challenge for robust isolation and detection of sources, including both cluster members and stars and galaxies in the foreground and background of the clusters.  To cover the large dynamic range of brightness and physical size of sources in the field, we adopt a hybrid source detection approach somewhat similar to the wavelet decomposition developed by \citet{livermore17}.  We first  create a master detection image, $D$, that is a combination of the ACS F814W and WFC3 F105W, F125W, F140W, and F160W mosaics weighted by the inverse variance maps of each.  Then we create two median-filtered versions of the detection image, $M_{16}$ and $M_{48}$, filtered on scales of 16 and 48 pixels (1\farcs6 and 4\farcs8), respectively.  The brightest, most extended sources are detected on the most smoothed image, $M_{48}$, with the \textsc{sep} source-detection software \citep{sep}.  Intermediate sources are detected on the $M_{16}-M_{48}$ filtered image, and a final detection is run on the $D-M_{16}$ image where all but the most compact features have been filtered out.  The final list of detected sources is the union of these three layers after removing duplicates between them.

\subsection{\textit{HST} photometry}
We extract aperture photometry within (circular) aperture diameters 0.36, 0.5, 0.7, 1.0, 1.2, 1.5, 3.0 arcsec at the positions derived in the source detection as described above.
We do not perform PSF-matching for any \textit{HST} filters for the aperture measurements. This PSF-matching approach is quite common in the literature (e.g. see \citealt{shipley18}), however, performing it on the images can result in substantial deleterious effects on the noise properties of the derived photometry at shorter wavelengths. In particular, faint and/or dropout sources would be most affected, where there is no signal to use in matching. In this work our scope is to focus on the faint, marginally detected objects. While the PSF - matching approach would result in, generally, more robust colors and photometric redshifts, we favor the approach here, primarily for simplicity and consistency across a wide variety of fields, with different noise properties, and defer tests on the aperture effects to ongoing work. Together with the derived photometry we also provide \textit{HST} mosaics for all filters, which can be utilised to perform the PSF - matched photometry extraction if necessary.
\par
To compute aperture corrections we have defined the ``total'' \textit{HST} flux density within an elliptical Kron aperture determined by \textsc{sep}, as in \textsc{SourceExtractor}. However, we do not impose the lower limit of 3.5 on \texttt{KRON\_RADIUS} typical with \textsc{SourceExtractor} as we find that in fact most derived values are actually lower than this threshold even for bright, well-measured sources. We do, however, impose a minimum circularized Kron aperture diameter of 0\farcs7, which is our favored ``color'' aperture. We calculate a correction for flux outside of the Kron aperture using the PSF curves of growth (i.e., explicitly valid only for point sources). The ``color'' aperture fluxes are therefore corrected by 1) \texttt{flux\_auto / flux\_aper} in the detection band and then 2) by the Kron aperture correction.

\begin{figure*}
\begin{center}
\includegraphics[width=.8\textwidth]{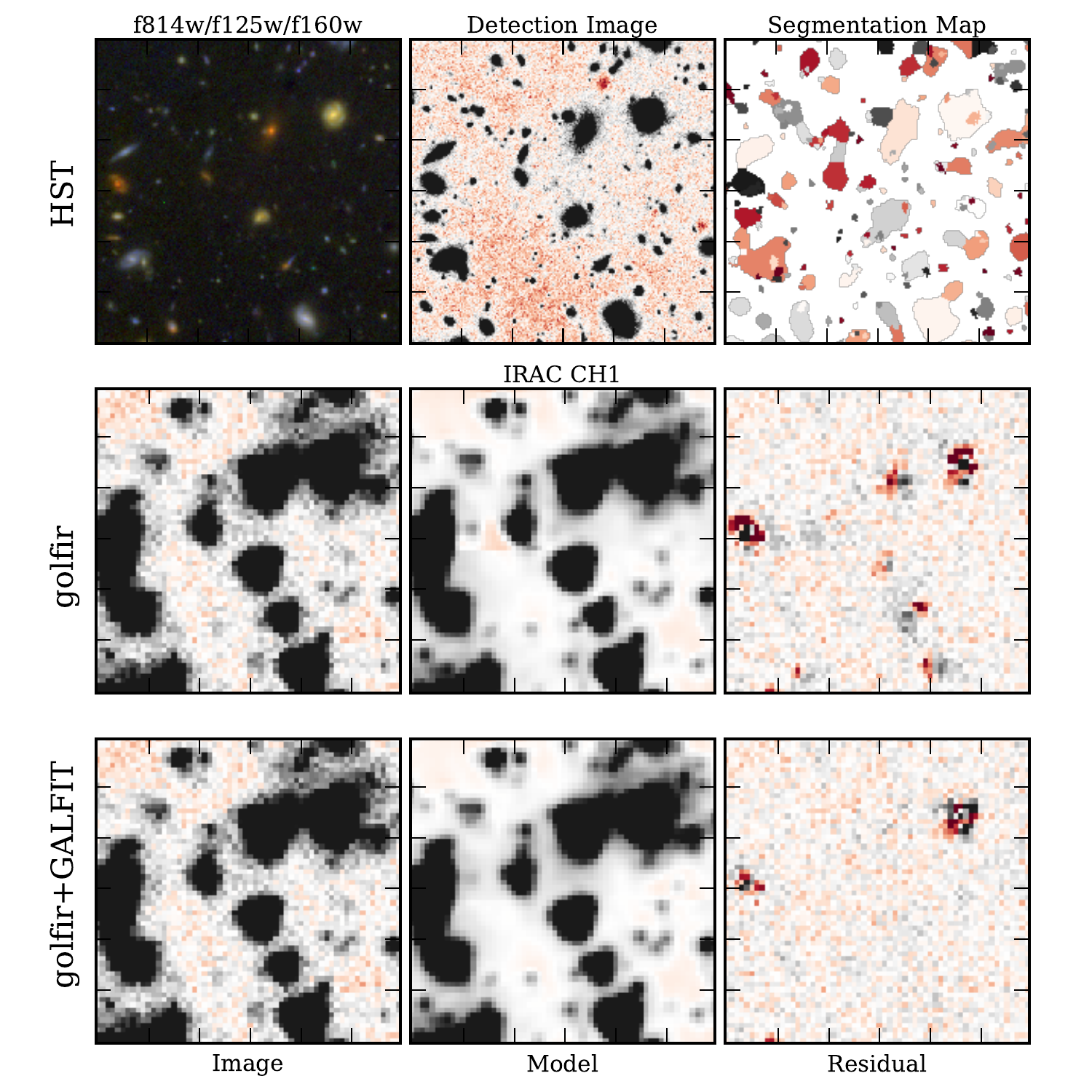}
\caption{Results from the modelling procedure on the relatively crowded Abell 370 cluster field. The cutouts are 30" across, and were selected to show a wide variety of sources on the same image, i.e. the ones modelled purely with \textsc{golfir}, and the galaxy model refined by using \textsc{GALFIT}. \textbf{Top:} The \textit{HST} RGB image created from the combination of F814W, F125W and F160W filters, the \textit{HST} detection image and segmentation map. \textbf{Middle:} The original IRAC 3.6 $\mu$m science image, the \textsc{golfir} model and residual images.
\textbf{Bottom:} Same as above, but now the model was additionally refined with \textsc{GALFIT}. The color scale has been adjusted to show the $3\sigma$ range for each image. Most notably, the final residual mosaic can also be used to extract additional IRAC photometry for sources without an \textit{HST} counterpart. The images are oriented with north up, and east to the left.}
\label{fig:resid}
\end{center}
\end{figure*}

\subsection{IRAC photometry}
Calculating the photometry from low-resolution data, particularly in crowded regions containing galaxy clusters can be a difficult technical challenge. In order to correctly extract the flux density in the redder bands, the significant differences that exist between the \textit{HST} data and the much lower resolution \textit{Spitzer}/IRAC image data must be taken into account. Primarily this concerns the problem of blending, wherein a standard aperture photometry approach normally used for high-resolution data would be inadequate. To tackle this issue in our work, we use the Great Observatories Legacy Fields IR Analysis Tools  or \textsc{golfir}, a set of tools developed to model \textit{Spitzer}/IRAC and MIPS images based on high resolution templates from existing \textit{HST} imaging, specifically in the context of the CHArGE data.\par
The method follows an approach utilized in similar lensing cluster catalog works (e.g. MOPHONGO; \citealt{Skelton14,shipley18} and T-PHOT; \citealt{merlin15,merlin16}), and relies on using a high-resolution prior. We create this prior by combining all the available \textit{HST} ACS/WFC and WFC3/IR filters, to produce a weighted mosaic, based on their corresponding inverse variance maps. We then use a convolution kernel to combine the detection image with the \textit{IRAC} PSF and produce the low-resolution templates. The original IRAC image is then divided into square patches, of user-defined size. These patches are also allowed to overlap, to allow for correct modelling of sources at the patch boundaries. For consistency we used a patch size of 1.2~arcmin with the 0.4~arcmin overlap for all the fields in our work. These parameters were chosen to achieve a balance between the quality of the final model image, and the available computational resources. 
\par
In order to improve the quality of extracted photometry we manually mask the brightest stars in a given field, prior to conducting the least square fit. We use the Gaia DR2 archive for the positions, and scale the size of the circular mask depending on the G-band magnitude of each star. We also mask all the pixels in the IRAC mosaic, for objects where the \textit{HST} catalog is brighter than AB = 15. This is done to avoid large residuals in the centers of bright stars that have not been manually masked. We additionally mask all pixels where the IRAC S/N is above 80.
\par
For the first model pass we generate IRAC model images for all objects in the \textit{HST} detection catalog, brighter than AB = 24, by using the aforementioned convolved \textit{HST} source cutouts, transformed into (position - dependent) IRAC PSFs. A least square fit of the low resolution IRAC cutouts is then performed to the real data, where the normalization acts as the only free parameter. We additionally derive any small residual shift between the reference \textit{HST} and target IRAC mosaics, by using the generated IRAC model and mosaic images.
\par
For the second model pass, we now focus on fainter \textit{HST} galaxies, with AB $<27$. IRAC models are then generated as before, with the least square normalizations now being adopted as the IRAC flux density measurement for each source. The diagonal of the covariance of the model design matrix is adopted as the photometric variance. Often there are clear systematic residuals in the fits for the brightest sources, which are likely due to a combination of 1) an imperfect transformation between the \textit{HST} and IRAC PSFs and 2) true morphological differences between F160W and the IRAC bands, for example the color gradients.  To improve the IRAC morphological model for those cases, we fit the IRAC images of all sources with total S/N $> 50$ directly with \textsc{GALFIT} \citep{peng02} assuming a single S\'ersic model \citep{sersic} and using the IRAC PSF.
We note that we are not interested in the \textsc{GALFIT} parameters, but rather in the best empirical description of each IRAC morphological component. We do not adopt the ``mag'' of the \textsc{GALFIT} fit, but rather refit the model normalizations and covariances as in the previous steps, now using the \textsc{GALFIT} model cutouts in place of the \textit{HST}-based models for the sources that have them. We show an example of the images processed with our pipeline in \autoref{fig:resid}, which also demonstrates the improvement in the model residuals after the \textsc{GALFIT} refinement.
\par
As the IRAC flux densities are based on morphological model fits, we consider them to be on the same "total" scale as the aperture-corrected \textit{HST} photometry. These are the IRAC flux densities that we will use for all our future data analysis. In addition to the model flux density fits, we also perform a simple aperture photometry measurement on our images, by using a $D$ = 3."0 apertures, similar to the approach taken in \citet{shipley18}. Using the IRAC PSF curves of growth we then correct the aperture flux densities into total flux density. The model, aperture and aperture corrected flux density measurements are all available in the final version of the catalog. In \autoref{tab:fields} we list all the names, coordinates and coverage areas of
all 33 cluster fields covered by ALCS. We would also like to note that for the Hubble Frontier Fields, we do not make a distinction between the parallels or the cluster, and rather treat the entire mosaic, and objects within it, to be contained within a single field.

\begin{deluxetable*}{ccccccc}
\tablecaption{\label{tab:fields}
	Cluster fields covered by ALCS}
\tablehead{%
Field	        	& RA 	& Dec	& Science Area$^\dagger$    & 3.6/4.5 $\mu$m Area   & ALMA (1.2 mm) Area & Redshift \\
    &                 (deg) & (deg) & (arcmin$^2$)  & (arcmin$^2$) & (arcmin$^2$) & $z_{spec}$   }
\startdata
\multicolumn{7}{c}{ALCS: Hubble Frontier Fields}\\
\hline \hline
Abell S1063 & 342.185 & -44.530 & 67.8 & 165.1 & 2.3 & 0.348 \\
Abell 370 & 39.970 & -1.577 & 74.4 & 197.9 & 3.3 & 0.375 \\
MACSJ0416.10-2403 & 64.037 & -24.075 & 68.1 & 191.3 & 2.3 & 0.396 \\
Abell 2744 & 3.588 & -30.397 & 64.7 & 172.3 & 2.7 & 0.308 \\
MACSJ1149.5+2223 &  177.401 & 22.399 & 37.6 & 190.5 & 2.6 & 0.543 \\
\hline
\multicolumn{7}{c}{ALCS: RELICS}\\
\hline \hline
RXCJ0032.1+1808  &	 8.046 & 18.130 & 11.6 & 54.0 & 6.4 & 0.396 \\
Abell 2537 & 347.093 & -2.192 & 11.5 & 55.0 & 2.0 & 0.297 \\
Abell 3192 &	59.721 & -29.929 & 11.5 & 55.0 & 4.0 & 0.425 \\
MACSJ0553.4-3342 &	88.346 & -33.708 & 11.5 & 48.2 & 6.9 & 0.430 \\
RXC J0600.1-2007 &	90.041 & -20.136 & 11.7 & 57.6 & 5.8 & 0.460 \\
RXC J0949.8+1707 &	147.462 & 17.121 & 11.6 & 55.4 & 2.6 & 0.383 \\
MACSJ0257.1-2325 &	44.293 & -23.437 & 13.3 & 57.4 & 1.7 & 0.505 \\
Abell 2163 & 243.951 & -6.127 & 22.0 & 97.7 & 1.5 & 0.203 \\
PLCK G171.9-40.7 & 48.237 & 8.372 & 11.5 & 39.7 & 3.8 & 0.270 \\
SMACSJ0723.3-7327 &	110.831 & -73.454 & 12.2 & 50.7 & 1.6 & 0.390 \\
MACSJ0035.4-2015 &	8.862 & -20.261 & 12.0 & 58.2 & 2.3 & 0.352 \\
MACSJ0417.5-1154 & 64.391 & -11.906 & 11.7 & 42.1 & 5.0 & 0.443 \\
MACSJ0159.8-0849 &	29.956 & -8.833 & 11.5 & 57.9 & 2.2 & 0.405 \\
ACT-CLJ0102-49151 &	15.750 & -49.273 & 22.9 & 106.6 & 4.4 & 0.870 \\
AbellS295 	& 41.381 & -53.040 & 11.8 & 48.5 & 3.2 & 0.300 \\
RXC J2211.7-0350 & 332.941 & -3.829 & 11.5 & 49.4 & 5.5 & 0.397\\
\hline
\multicolumn{7}{c}{ALCS: CLASH}\\
\hline \hline
Abell 383 & 42.014 & -3.529 & 16.8 & 62.0 & 0.8 & 0.187 \\
MACS1206.2-0847 & 181.551 & -8.801 & 13.1 & 55.0 & 2.0 & 0.440 \\
MACS1423.8+2404 & 215.949 & 24.078 & 15.2 & 59.9 & 1.2 & 0.545 \\
MACS1931.8-2635 & 292.957 & -26.576 & 13.1 & 49.1 & 1.8 & 0.352 \\
RXJ 1347-1145 & 206.877 & -11.753 & 12.6 & 46.8 & 2.5 & 0.451 \\
MACS1311.0-0310 & 197.757 & -3.178 & 13.2 & 52.8 & 0.9 & 0.494 \\
MACS1115.9+0129 & 168.967 & 1.499 & 13.2 & 53.0 & 1.0 & 0.352 \\
MACS0429.6-0253 & 67.400 & -2.886 & 13.2 & 54.2 & 0.7 & 0.399 \\
RXJ2129.7+0005 & 322.416 & 0.089 & 18.4 & 75.5 & 0.5 & 0.234 \\ 
MACS0329.7-0211 & 52.424 & -2.197 & 12.8 & 48.7 & 2.1 & 0.450 \\
MACS2129.4-0741 & 322.359 & -7.691 & 14.7 & 55.8 & 1.7 & 0.570 \\
Abell 209 & 22.969 & -13.611 & 15.1 & 54.1 & 0.7 & 0.206 \\
\enddata
\begin{tablenotes}
$\dagger$ \footnotesize{The ''Science Area'' corresponds to the coverage of the detection band, as defined in \autoref{sec:phot}}
\end{tablenotes}
\end{deluxetable*}

\section{Catalog} \label{sec:cat}

\subsection{ALMA Photometry}
\label{sec:alma_phot}
\begin{figure*}
\begin{center}
\includegraphics[width=.95\textwidth]{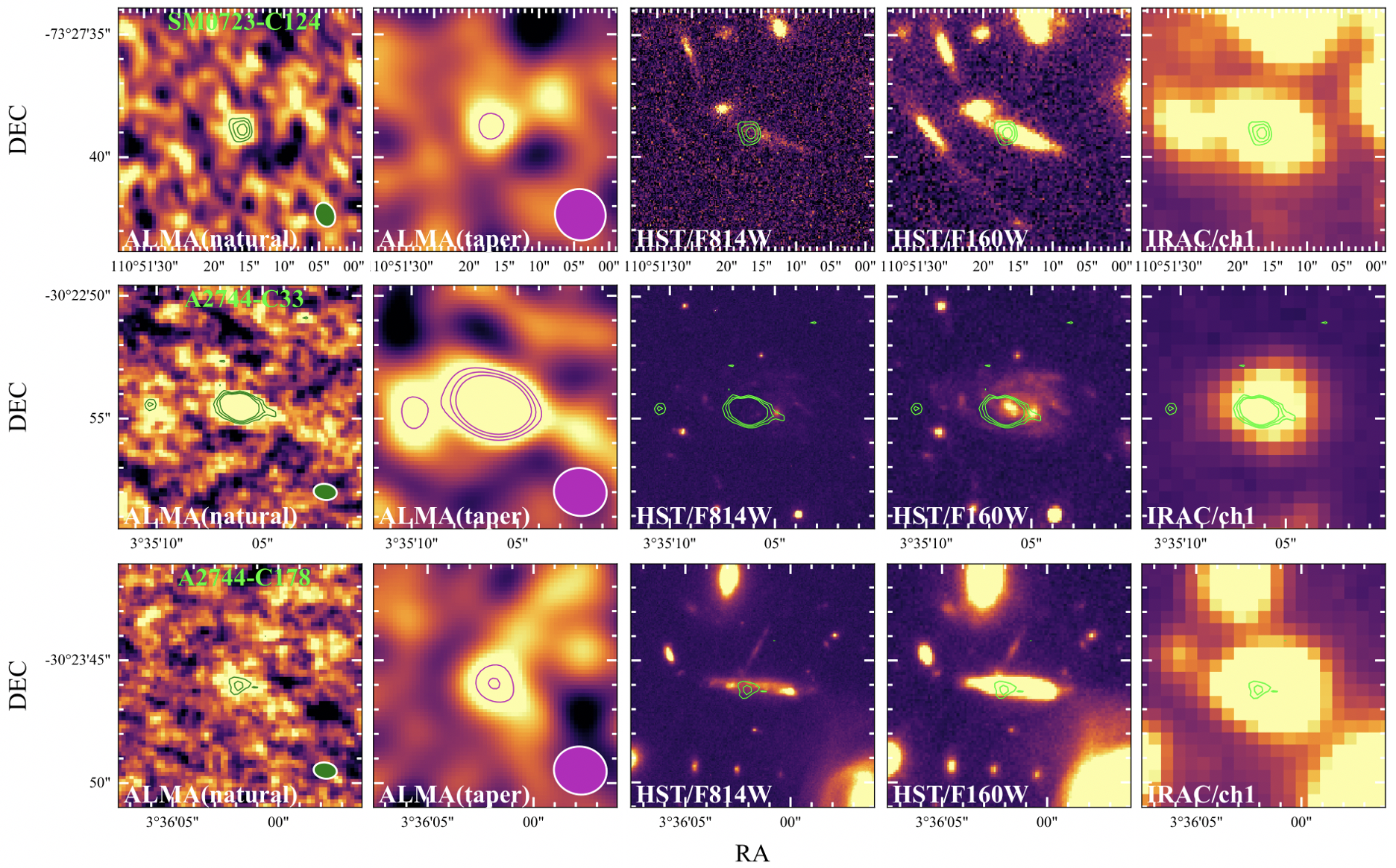}
\caption{Image stamps displaying natural and tapered ALMA maps, alongside with a cutout from $HST$/ACS $HST$/WFC3 and IRAC for three sources in the SMACS0723 and Abell 2744 fields. On top of each image we overlay 1,2 and 3 $\sigma$ contours. Green and purple shapes at the bottom of ALMA cutouts represent the size of the beam. Each panel is 10" across. The remainder of the stamps for each source will be presented in the upcoming work (Fujimoto et al. 2022; in prep) and are also available in the same repository as our catalogs.}
\label{fig:alma_gal}
\end{center}
\end{figure*}

To further complement our photometric catalog, we have included an additional data entry containing either the measurement or the upper limit on the ALMA flux density. To do this, we start by cross-matching our objects with the ALMA continuum catalog (Fujimoto et al. 2022; in prep). This catalog contains 180 sources, which have been selected with a S/N cut $>4$. The total flux densities are computed as a peak count, after primary beam correction in the tapered map [mJy/beam]. If no peak is identified in the tapered map within a radius of 1.0 arcsec, the pixel count at the position of the source is used instead. We do not match these sources automatically, as this approach is inadequate for nearby or highly magnified sources with complex image plane morphology. In addition to the the astrometry difference, the ALMA beam size might result in erroneous counterpart assignment.
Instead, we manually examine ALMA contours for each detected object, overlaid on the $HST$ and $Spitzer$ cutouts and use them to assign counterparts. We feel that for crowded fields with complex lens geometry, this is the most secure approach. In \autoref{fig:alma_gal} we present a series of example stamps for three ALMA detected objects in SMACS0723 and Abell 2744 fields. The rest of the 180 cutouts will be presented in Fujimoto et al. 2022 (in prep.) and are also available online\footnote{\url{https://github.com/dawn-cph/alcs-clusters/tree/master/v1.0/alma_cutouts}}. In total, we find 145/180 matches during this procedure. These are flagged with $\texttt{alma\_coverage}=2$ in the catalog. From the 35 objects that do not have a match to our catalog: 14 are
``\textit{HST}-dark'' with clear detections in both IRAC channels; 8 are not visible in either $HST$ or IRAC; 10 fall outside of the $HST$ area, but have clear IRAC detections; 3 are too close to a star, or some other image artifact. The $HST$ and IRAC "dark" objects are cataloged separately.

Apart from the aforementioned 145, the remaining objects in our catalog do not have reliably measured ALMA counterparts, however the ALMA continuum map at 1.2 mm still overlaps significantly with our mosaics, and as such it is still possible to extract the ALMA upper limits at the positions of our sources. We begin the extraction of upper limits by isolating a sub-sample of sources in our catalogue that do not have an ALMA counterpart in the continuum catalogue, but still fall within an ALMA map. The objects that fall outside of ALMA coverage have been flagged with $\texttt{alma\_coverage}=0$. Within each field, we have used an approach similar to the original flux density extraction of Fujimoto et al. (2022; in prep) and measured the flux density from the central pixel on the ALMA map. As the uncertainty on the measured flux density we adopted the noise level of the entire map (as described in Fujimoto et al. 2022, in prep), calibrated to the area of the beam. This procedure results in 30,586 galaxies with a 1.2 mm upper limit, which is equivalent to $\sim 14 \%$ of the combined catalog source count of 217,958. While these are only upper limits, the addition of an extra constraint in the FIR will significantly enhance the quality of panchromatic SED fitting. For example, these measurements can help to compute the upper limits on the dust mass ($M_{\rm dust}$), for objects up to $z\sim 7$, and the infrared luminosity $L_{\rm IR}$ above $z \sim 9$. This is due to the fact that the 1.2 mm stops sampling the rest-frame continuum above 150 $\mu$m at $z\sim7$, which is required for robust calculations of the dust mass (e.g. see discussion in \citealt{berta16} and \citealt{kokorev21}). In return, however, the 80-120 $\mu$m regime will then become available for 1.2 mm photometry at $z=9$, which would allow to sample the peak of the FIR emission and allow to impose tighter constraints on both total infrared luminosity $L_{\rm IR}$ and the dust temperature - $T_{\rm dust}$. This naturally depends on how dusty, high-$z$ galaxies actually are, and whether the commonly used optically thin dust approximation will apply to them.

\subsection{Catalog Description and Flags}
We provide separate multi-wavelength photometric catalogs for each of the 33 clusters covered by ALCS. Combined, the catalogs contain aperture and total flux density measurements for 218,000 sources. We list the description of the relevant columns of the catalog in \autoref{tab:catcols}. All flux densities and associated uncertainties are in units of $\mu$Jy, unless specified otherwise. For the \textit{HST} data we include the aperture photometry, measured within $D=$0\farcs7, and the associated total flux densities. The IRAC measurements are provided as the \textsc{golfir} modeled IRAC photometry, as well as aperture (and aperture corrected) flux densities, measured within $D=$3\farcs0. The cross-matched ALMA photometry is provided in units of $\mu$Jy/beam. We also include an ALMA source flag column - \texttt{alma\_coverage}, where 2 - indicates a detection, 1 - an upper limit and 0 - a lack of coverage. To help discern between the high and low quality photometry for sources in the catalog, we have added a number of flags which allow to uniformly select reliable samples of objects. For each source we compute how many pixels in a 3x3 square around the center have been masked or fall outside the IRAC mosaic. This number is given in the \texttt{n$\_$masked} column. If the central pixel itself is masked or missing, we flag that source with \texttt{bad$\_$phot} = 1.

\begin{deluxetable}{ccc}[h]
\tabcolsep=0.15mm
\tablecaption{\label{tab:catcols}
	Description of the Relevant Photometric Catalog Columns}
\tablehead{%
Column Name	  & Units      	& Description}
\startdata
id & - & Object ID\\
ra & deg & Right Ascension \\
dec & deg & Declination \\
\{\texttt{filt}\}\_flux\_aper & $\mu$Jy & \textit{HST} $D$=0\farcs7 aperture flux  \\
\{\texttt{filt}\}\_err\_aper & $\mu$Jy & Uncertainty \\
\{\texttt{filt}\}\_flux & $\mu$Jy & Total \textit{HST} flux density  \\
\{\texttt{filt}\}\_err & $\mu$Jy & Uncertainty \\
irac\_\{\texttt{ch}\}\_flux & $\mu$Jy & IRAC model flux\\
irac\_\{\texttt{ch}\}\_err & $\mu$Jy & Uncertainty\\
irac\_\{\texttt{ch}\}\_flux\_aper & $\mu$Jy & IRAC $D$=3.0" aperture flux\\
irac\_\{\texttt{ch}\}\_err\_aper & $\mu$Jy & Uncertainty\\
alma\_coverage & - & ALMA coverage flag \\
alma\_flux & $\mu$Jy & ALMA flux at 1.2 mm \\
alma\_err & $\mu$Jy &	Uncertainty\\
\enddata
\end{deluxetable}

\subsection{Comparison to the existing catalogs} \label{sec:qual}
Provided that there are existing catalogs in the lensed fields included in ALCS, it is also useful to perform a comparison of our catalog and those presented in the literature. These include the publicly available photometric catalogs in the Hubble Frontier (HFF-DeepSpace \citet{shipley18}; ASTRODEEP \citet{astrodeep1,astrodeep2}) and CLASH fields \citep{molino17}. Series of diagnostic plots, including broad-band photometry, color-color diagrams and number counts are fully presented and discussed in the \autoref{sec:appendix_1}.

\section{Galaxy Properties} \label{sec:param}
To derive the photometric redshifts, rest-frame color, dust attenuation, and stellar population parameters we have used the updated Python version of \textsc{EAZY} \citep{brammer08}. \textsc{EAZY} is a photometric spectral energy distribution fitting code which is optimised to fit non-negative linear combinations of basis-set templates, rather than devising a solution from vast libraries of pre-compiled models. These templates are extracted from the Flexible Stellar Populations Synthesis (FSPS; \citealt{conroy09}) models, and then reduced to a set of 12, which are able to reproduce a much larger library, spanning a variety of dust attenuation, ages, mass-to-light ratios and star-formation history (SFH) properties (see \citealt{brammer08} and \citealt{br07}). We have listed the properties for these individual templates in \autoref{tab:template-params}. For each object in our multi-wavelength catalogue, \textsc{EAZY} integrates redshifted templates through a set of filters corresponding to our observed bands, and then finds the best combination for a given set of flux densities and associated uncertainties. 
\par
To improve the quality of the photometric redshift estimate, \textsc{EAZY} also implements a ``template error function'', which is used to account any other miscellaneous uncertainties related to short-lived and otherwise unusual stellar populations, and the emission lines coupled to the properties of the ISM. In our analysis we have used a default template error function value of 0.2. To calculate the photometric redshifts and physical parameters with \textsc{EAZY} we have used total \textit{HST} flux densities, calculated from the $D=$0\farcs7 aperture measurement, and the \textsc{golfir} IRAC models. These data are contained in "\{\texttt{filt}\}\_flux" and "irac\_\{\texttt{ch}\}\_flux" columns, respectively.
\par
From the best fit \textsc{EAZY} SEDs we derive the stellar population properties which include, but are not limited to the SFR, $A_{V}$, $M_*$ and rest-frame colors.

\subsection{Spectroscopic Redshift Catalogs}
The ALCS fields have been covered by an wide range of spectroscopic surveys. We have examined the literature and compiled all of the available spectroscopic redshifts in order to assess the quality of our photometric redshifts. During this process we select only the sources with robust redshift constraints, and choose the most recent source, if a galaxy is present in multiple catalogs. We have cross-matched all the available spectroscopic redshifts with objects in our catalogue, using a 1\farcs5 matching radius. These spectroscopic redshifts are included in our main catalogue, with a separate column providing the source where available. 

For Hubble Frontier Fields we have used spectroscopic redshifts already compiled in \citealt{shipley18}, this includes Grism Lens-Amplified Survey from Space (GLASS) \citep{treu15,schmidt14}, and the spectral data presented in \citealt{smith09,owers11,ebeling14,jauzac14,richard14,balestra16,caminha16,diego16,grillo16,karman2016,limousin16,lagattuta17,mahler18}. Where appropriate, we also have updated the spectroscopic redshifts with the most recent results from \citealt{richard21}. In total we recover 5,055 matches with our catalogue, in all five HFF fields, including the parallels. Please note however, that the number of spectroscopic redshifts in the parallel fields is severely limited. We include them in our final comparison for the record.
\par
For CLASH we have compiled the most recent data from GLASS \citep{treu15,schmidt14}, the results from \citealt{caminha19} and \citealt{richard21}. In total there are 2,090 matches with our catalogue, spanning ten out of twelve CLASH fields. Finally, for the RELICS spectroscopic redshifts, we have used the data from \citet{richard21} for RXC0600, and a compilation of spec-$z$ data presented in Guerrero et al. (2022, in prep). In total we find 351 matches with our catalogs. The full list of fields and matches is outlined in \autoref{tab:zspec}.

\subsection{Photometric Redshift Accuracy}
\begin{figure}
\begin{center}
\includegraphics[width=.45\textwidth]{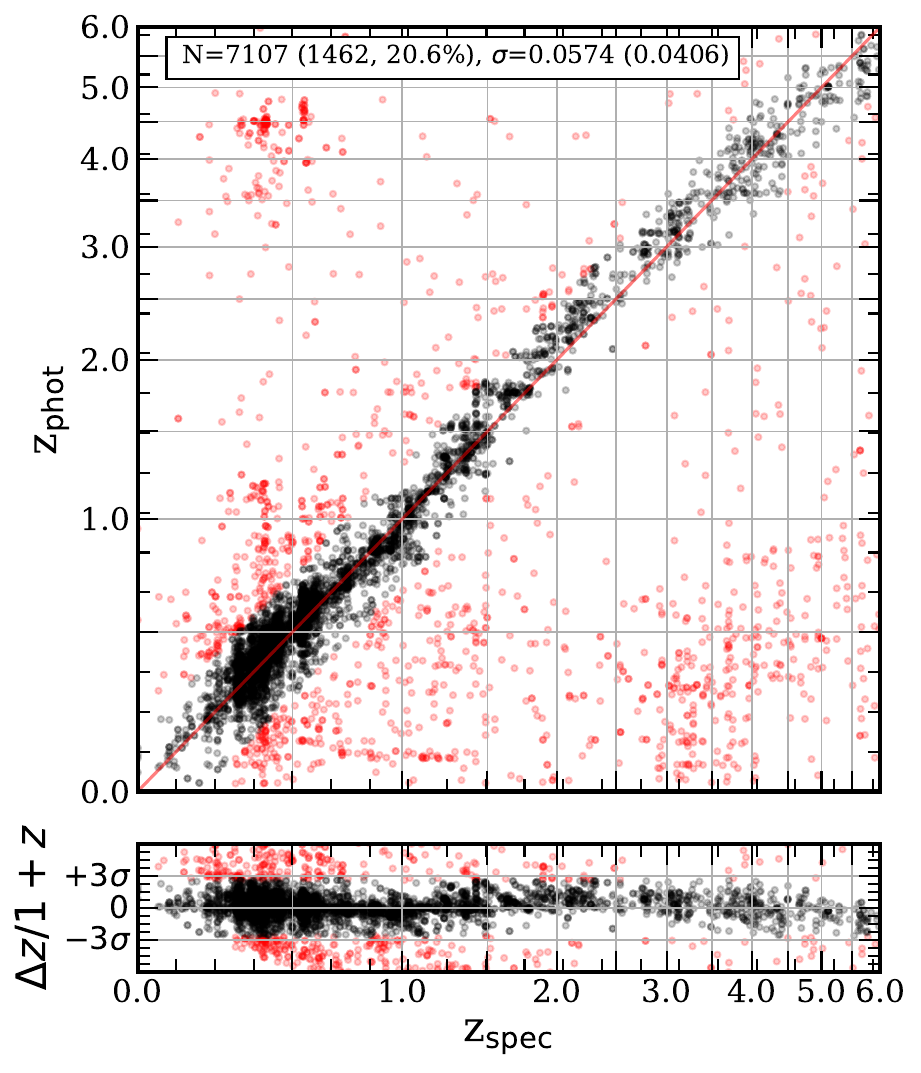}
\caption{The comparison between photometric redshifts derived by \textsc{EAZY} and the spectroscopic redshift from the literature. We have carried out the comparison in all fields where the match was found, in total these include all five Frontier Fields, ten CLASH fields, and nine RELICS field. Black and red circles denote galaxies below and above catastrophic limit of 0.15, respectively.}
\label{fig:zspec}
\end{center}
\end{figure}

 To quantify the precision of our photo-$z$ estimate we used the normalised median absolute deviation (NMAD, \citealt{hoaglin83}), defined as:
\begin{equation}
\sigma_{\rm NMAD}=1.48 \times \mathrm{median} \left(\frac{|\Delta z - \mathrm{median}(\Delta z)| }{1+z_{\rm spec}}\right).
\end{equation}
This method in commonly used in the literature (e.g. see \citealt{shipley18,Skelton14}), allowing for a quick and unbiased comparison of redshift quality between different catalogs, and is also less sensitive to outliers as described in \citealt{brammer08}. The outlier fraction $\eta$ is given by $|\Delta z|$/(1+$z_{\rm spec})>0.15$, following the methodology described in \citealt{hildebrandt12}. \\
In total we have carried out the comparison for 7,107 matched objects, excluding catastrophic outliers, spanning 24 fields, shown in \autoref{fig:zspec}. We find that our redshift accuracy is generally good, with a $\sigma_{\rm NMAD}$ of 0.0406, and 20.6 $\%$ of catastrophic failures.

For both the $z_{\rm phot}-z_{\rm phot}$ and $z_{\rm phot}-z_{\rm spec}$ comparisons, we note the existence of over-densities located either at $z\sim 1$ or $z\sim 4$, for which $\Delta z\sim3$. This is where a vast majority of our catastrophic outliers are located. These redshift discrepancies are caused by the mis-identification between the Lyman (912 \AA) and the Balmer (3644 \AA) breaks, as well as the 4000 \AA\, break, in the fitted SEDs. The manifestation of the break confusion is a consequence of degenerate behaviour of templates, when faced with either sparse or very faint photometry, and is particularly prominent when we compare large samples of photometric redshifts between different catalogs.

\subsection{Gravitational Lensing Magnification}
\begin{figure*}
\begin{center}
\includegraphics[width=.85\textwidth]{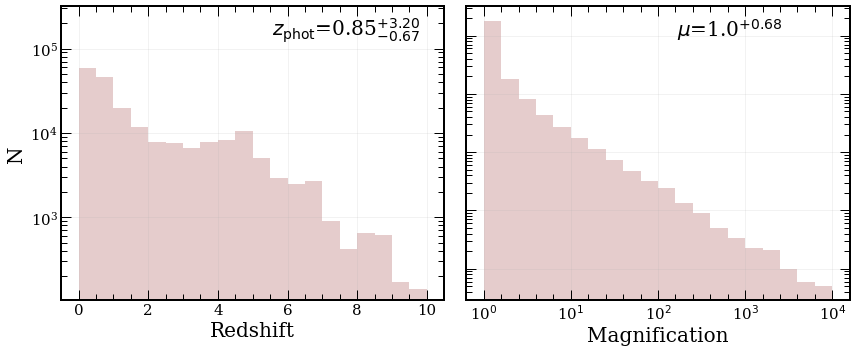}
\caption{The log-scaled distribution of best-fit photometric redshifts (left) and computed magnification factors (right) for the full catalog. The median and the 68 \% confidence interval for each parameter are shown on each plot.}
\label{fig:z_mu_distr}
\end{center}
\end{figure*}

\begin{figure}
\begin{center}
\includegraphics[width=.49\textwidth]{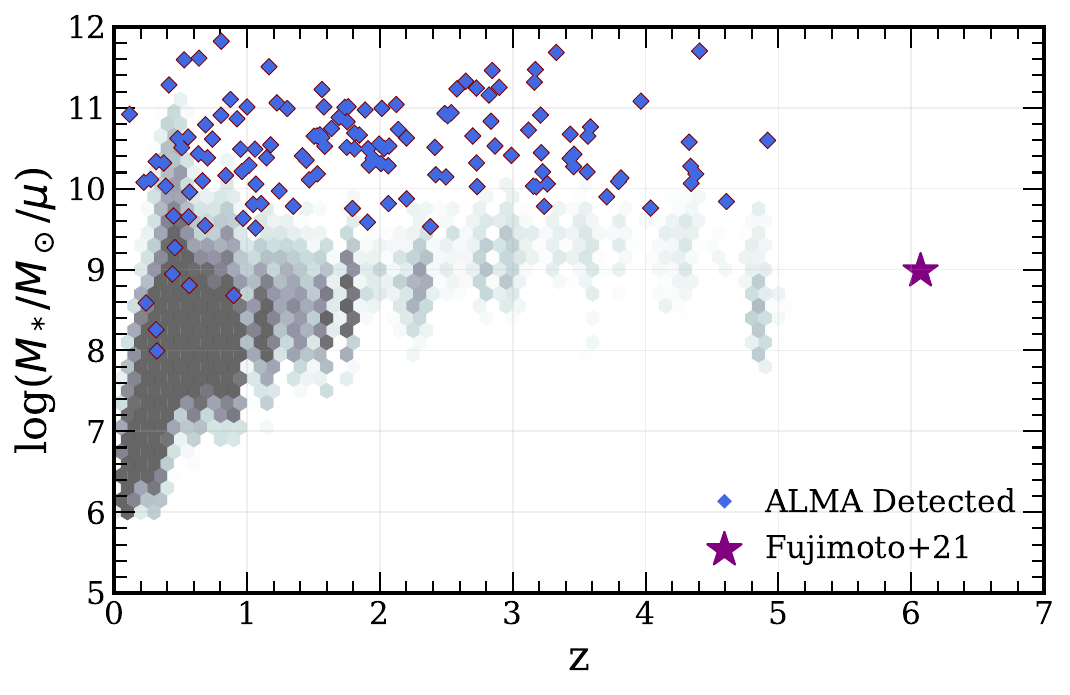}
\caption{A density plot of \textsc{EAZY}-derived de-magnified stellar mass, as a function of redshift. ALMA - detected galaxies are shown as blue diamonds. A bright, multiply lensed $z=6.07$ source from \citet{fujimoto21} is shown as a purple star.}
\label{fig:mstar_alma}
\end{center}
\end{figure}

For objects within the cluster fields we compute and provide the lensing magnification factor ($\mu$), which is based on the RA,DEC coordinate of the source in the detection band (i.e. the peak flux density coordinate) and its redshift. Although a vast majority of sources in a given field only have a $z_{\rm phot}$ estimate, we use a $z_{\rm spec}$ where possible. Following the methodology described in \citet{sun22}, we use the Zitrin-NFW lens models \citep{zitrin13,zitrin15} for the HFF and CLASH clusters, and GLAFIC models \citep{oguri10,okabe20} for RELICS. These models consist of the mass surface density ($\kappa$) and weak lensing shear ($\gamma$) maps. We then compute the magnification by using:
\begin{equation}
\mu=\frac{1}{(1-\kappa\,\beta)^2-(\gamma\,\beta)^2},
\end{equation}
where $\beta$ is the lensing depth, defined as $\beta={D_{\rm ls}}/{D_{\rm s}}$, with $D_{\rm ls}$ being the angular diameter distance between the lens and the source, and $D_{\rm s}$ is the angular diameter distance to the source. Similarly to \citealt{rawle16}, if the source redshift $z_{\rm s}$ is below or within the cluster redshift $z_{\rm cl}$, namely $z_{\rm s}\leq z_{\rm cl}+0.1$, we set the magnification to unity. Sources that fall outside of the magnification maps for a given field, are not expected to be significantly affected by gravitational lensing. For these, we have set $\mu=1$. In \autoref{fig:z_mu_distr} we present the distributions of best-fit photometric redshifts and magnification values for all objects in our catalog.
Please note that we do not apply these lensing corrections to any flux densities listed in our catalog. \\
We present the distribution of the de-magnified $M_*$ with redshift in \autoref{fig:mstar_alma}. In addition to that, we also highlight the ALMA detected ALCS sources, including the \citet{fujimoto21} object at $z=6.07$. We note that ALMA detected objects are more massive when compared to the other galaxies in the catalog. A similar trend has been observed in both in $M_{1500}$ vs $z$, and $M*$ vs $z$ relations in \citet{dunlop17}.

\subsection{Red Sequence}
In lieu of secure spectroscopic redshifts, a definitive assessment of cluster membership for galaxies in our catalog can not be achieved.
While the $z_{\rm phot}$ probability density functions - $p(z)$, and the comparison to $z_{\rm spec}$ via $\sigma_{\rm NMAD}$ can provide us with important hints regarding the robustness of photometric redshifts, an additional quality test is required. \par
We can exploit the observational fact which dictates that the majority of early-type galaxies (ETGs), that consist in massive clusters, display a tight color-magnitude correlation. This relation has been referred to as the Cluster Red Sequence, or just the red sequence \citep{gladders00}. This color - magnitude relation was shown to hold from cluster to cluster \citep{lopezcruz04} and also displays a relatively small scatter \citep{bower92}. The technique provides an independent redshift constraint for galaxies contained within clusters, with the ETG overdensities being exploited as a cluster marker for quite some time in the literature \citep{lopezcruz97,gladders98,kaiser98,yee99,lubin00,lopezcruz04}. \par
\begin{figure*}
\begin{center}
\includegraphics[width=.8\textwidth]{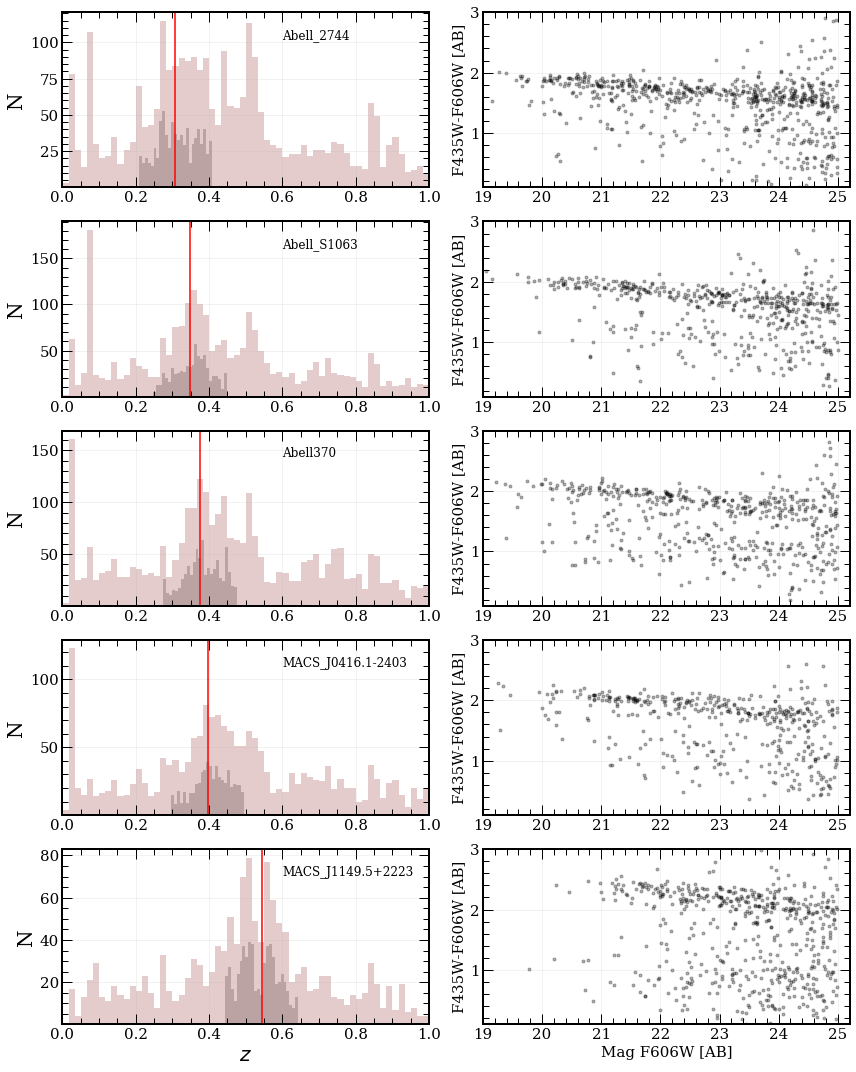}
\caption{Red sequence diagnostic for galaxies within Hubble Frontier Fields. \textbf{Left:} A histogram of all objects in the catalog in the $0<z<1$ range (red) and an isolated sample of galaxies which we consider to belong to the cluster (grey) -  $|z-z_{\rm clu}| < 0.1$ and F606W$<25$ AB mag. The vertical red line represents the redshift of the cluster. \textbf{Right:} Color - magnitude diagnostics of the red sequence with the F435W and F606W filters.}
\label{fig:hff_red_seq}
\end{center}
\end{figure*}

Effectively the cluster galaxies which should lie on the red sequence can be isolated from the field objects with only two filters that cover the 4000 \AA\, break. For our dataset this can be achieved through a combination of the F435W and F606W \textit{HST}/ACS filters. In \autoref{fig:hff_red_seq} we present a red sequence diagnostic plot for the HFF galaxies that are considered to be within clusters based on their photometric redshift ($|z_{\rm phot}-z_{\rm cluster}|\leq0.1$) and are also bright ($m_{\rm F606W}<25$). By examining the distribution of galaxies as a function of redshift we note the presence of distinct density peaks located in proximity to the cluster redshift. The color - magnitude diagnostic plot also reveals that galaxies which we have selected to be a part of the cluster based on the $z_{\rm phot}$ tend to form a distinct color sequence on the diagnostic plot.
From the combination of both $\sigma_{\rm NMAD}$ and red sequence diagnostics, we conclude that the recovered photometric redshifts, and the stellar population parameters derived based on them, are robust both for the field and cluster galaxies in our catalogs.

\subsection{Rest-Frame Color Galaxy Classification}
Comparison of galaxies at different redshifts often requires rest-frame, rather than observed frame, colors to be used. For each galaxy these are determined by assuming the best-fit \textsc{EAZY} template and its redshift, in order to calculate the rest-frame flux density for a set of filters. The rest-frame colors are then computed by integrating a transmission curve for a given filter through the best fit template, as described in \citealt{brammer11}. In our final catalogs we provide the rest-frame flux densities for the most commonly used filters ($GALEX$ $NUV$, COSMOS $r$, Johnson $U$, $B$, $V$, and $J$).

\begin{figure*}
\begin{center}
\includegraphics[width=.9\textwidth]{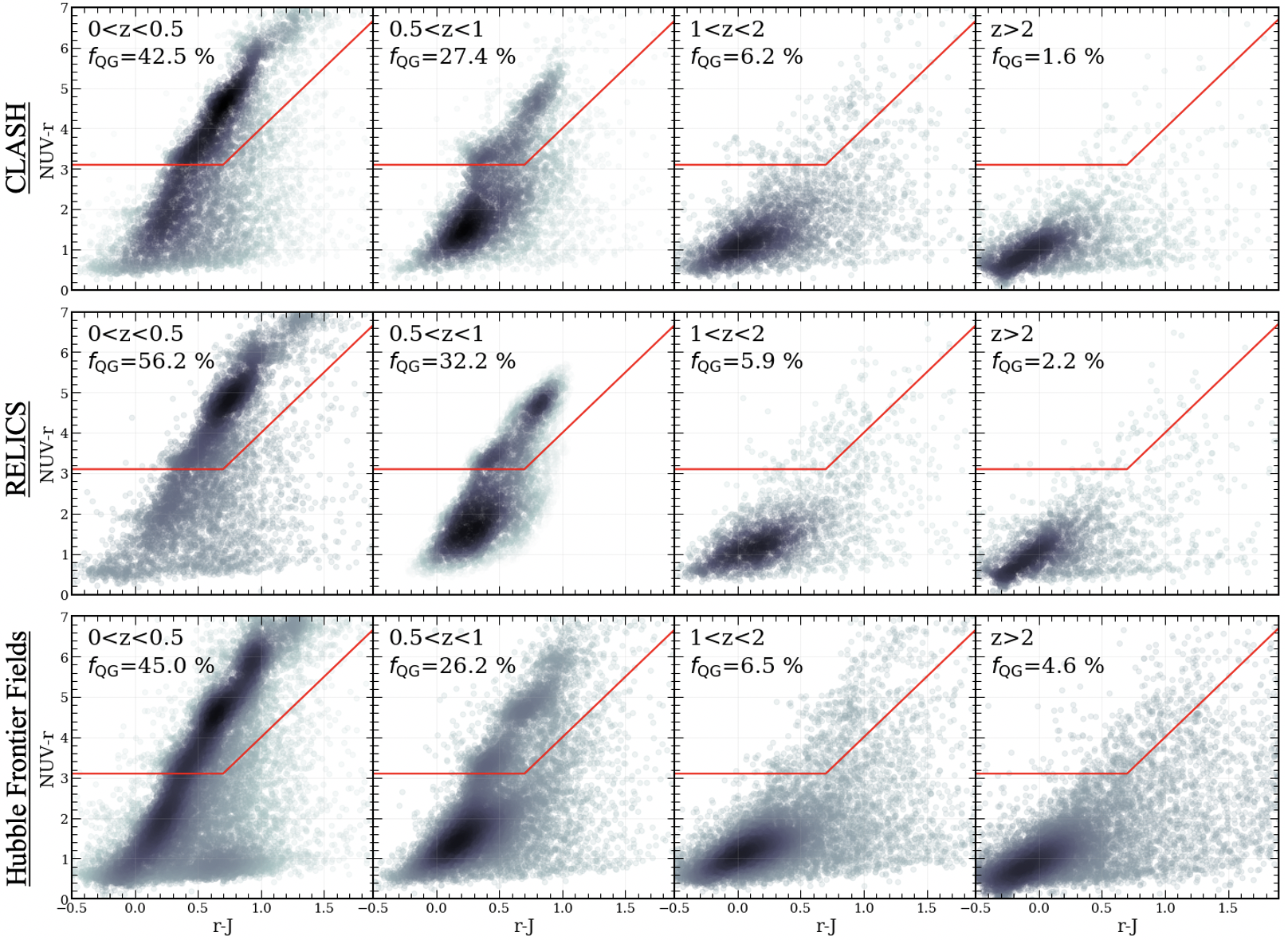}
\caption{Classification of galaxies for all ALCS fields by using the $NUV-r$ and $r-J$ rest frame colors. The $NUV-r-J$ galaxy selection prescription from \citet{ilbert13} is shown as a solid red line. The fraction of quiescent galaxies (QGs) is displayed within each plot. Here we have limited our selection to only include galaxies with \texttt{nusefilt}$\geq5$ and a S/N $> 5$ in the $HST$ f160w filter.}
\label{fig:nuvrj}
\end{center}
\end{figure*}

The rest-frame flux densities can be used to assess the galaxy populations in each field by using the color-color analysis. Multiple previous studies have devised a variety of techniques to classify galaxies based on the broad-band photometry. One such prescription utilises the $U-V$ and $V-J$ rest-frame colors \citep{labbe05,wuyts07,williams09} to separate galaxies into quiescent, star-forming. Quiescent galaxies with low levels of star formation are red in the $U-V$ regime and are easily distinguished from the similarly red (in $U - V$), dusty, star-forming galaxies, with the $V-J$ color.
An alternative method, using the $NUV-r-J$ colors instead, has been proposed by \citet{ilbert13} and \citet{arnouts13}. While the $UVJ$ selection is the most commonly used approach in the literature, the $NUVrJ$ method has some key advantages. The shorter wavelength $NUV$ band is more sensitive both to the dust attenuation and emission from young stellar populations than the $U$ band. Although the amount of quiescent galaxies at $z>2$ is limited (see e.g. \citealt{ilbert13,muzzin13,davidzon2017}), the rest-frame $NUV$ will still be covered by the optical photometry at that redshift regime, which is no longer the case for the rest frame $U$ band. This, in return, would make the $NUVrJ$ selection technique more reliable at higher redshifts. In our paper we adopt the $NUVrJ$ selection, however the $U$ and $V$ rest frame flux densities are also provided in our final catalog.
We display our selection in \autoref{fig:nuvrj}, where the quiescent galaxies tend to be located in the upper-left corner of the diagram, with the boundaries defined in \citet{ilbert13}:
\begin{equation*}
NUV-r=
 \begin{cases}
    3\, (r-J)+1       & \quad \text{for} \quad r-J>0.7 \\
    3.1    & \quad \text{for}  \quad r-J<0.7.
 \end{cases}
\end{equation*}

When comparing the $UVJ$ and $NUVrJ$ color classification methods, $\sim 70 \%$ of objects selected with $NUVrJ$ overlap with $UVJ$ quiescent candidates, varying slightly with the depth of our field of choice, and therefore its redshift distribution.
Overall we find that 17 \% of objects in HFF, 23 \%  in RLC and 17 \% in CLS are located in the upper quadrant of the $NUVrJ$ diagram, and can thus be classified as quiescent. The ALCS covers cluster fields, where the star formation is generally expected to be suppressed, especially at low redshift (see e.g. \citealt{boselli16} and references therein). Therefore it is not at all surprising for us to recover high fractions of quiescent galaxies.

\begin{table}
\caption{Templates used for the \textsc{EAZY} fit.}
\centering
\footnotesize
\label{tab:template-params}
\tabcolsep=0.1mm
\begin{tabular}{cc}
\hline\hline
Parameter & Value \\
\hline
\multicolumn{2}{c}{Optical emission: \citealt{brammer08}$^a$}\\
\hline
$A_V$$^b$ & [0.6,0.12,0.19,0.29,1.05,2.68,\\
& 0.11,0.36,0.98,1.54,1.97,2.96]\\
\hline
$M/L_V$ & [0.38,0.76,1.68,4.01,6.45,44.48,\\
& 0.12,0.21,0.33,0.64,1.57,4.00]\\
\hline
log$_{10}$(sSFR) & [-10.75,-11.37,-11.90,-12.53,-12.05,\\
& -12.47,-8.37,-8.60,-8.50,-8.57,-8.93,-8.90]\\
\hline
\hline
\end{tabular}
\begin{tablenotes}
\item[a] \footnotesize{Please refer to \citealt{brammer08} for a more detailed description of the creation and selection of these basis set templates. See \citealt{br07} for a methodology regarding the SFH.}

\item[b] \footnotesize{\citet{calzetti00} \hfill}
\end{tablenotes}
\end{table}

\begin{deluxetable}{cc}
\tabcolsep=1.cm
\tablecaption{\label{tab:zspec}
Sources of Spectroscopic Redshift.}
\tablehead{%
Field &  Matches \\
& (\# of Galaxies) }
\startdata
\multicolumn{2}{c}{HFF: cluster \& parallel}\\
\hline
Abell 2744 & 736  \\
Abell 370 & 793 \\
MACSJ0416  & 1852 \\
MACSJ1149 & 1173\\
Abell S1063 & 501 \\
\hline
\multicolumn{2}{c}{CLASH}\\
\hline
MACS0329 & 129  \\
MACS0429 & 129 \\
MACS1115  & 66 \\
MACS1206 & 424\\
MACS1311 & 68\\
RXJ\_1347 & 511\\
MACS1423 & 154\\
MACS1931 & 138\\
MACS2129 & 322\\
RXJ2129 & 149\\
\hline
\multicolumn{2}{c}{RELICS}\\
\hline
RXCJ0600 & 73\\
ACTCLJ0102-49151 & 76\\
AbellS295 & 15\\
Abell 2163 & 44\\
Abell 2537 & 37\\
MACSJ0159.8-0849 & 9\\
MACSJ0257.1-2325 & 11\\
MACSJ0417.5-1154 & 25\\
MACSJ0553.4-3342 & 61\\
\enddata
\end{deluxetable}

\section{Conclusions} \label{sec:conc}
This paper describes the creation of \textit{HST}+IRAC photometric and galaxy property catalog within the 33 lensed cluster fields, covered by the 134 arcmin$^{2}$ ALCS survey. The mosaics and catalogs cover a combined area of $\sim690$ arcmin$^2$, in 33 ALCS fields, which include 5 Hubble Frontier, 16 RELICS, and 12 CLASH fields.
The final catalog numbers roughly 218,000 sources, which are covered by at most 12 \textit{HST}/ACS,UVIS and WFC3 bands, plus the additional IRAC photometry at 3.6 and 4.5 $\mu$m. To process these data we have
reprocessed and recombined all the available archival \textit{HST} exposures, now combined into a single CHArGE dataset, as well as all available IRAC data covering the same fields. Each image has been aligned to the same highly precise Gaia DR2 reference frame, ensuring a robust internal alignment of the \textit{HST} and IRAC images for matched-aperture photometry, with the final absolute astrometric precision generally being $<100$ mas. 
\par
In our analysis, we have applied a consistent methodology in order to compute multi-wavelength photometry across all 33 fields. The \textsc{sep} software \citep{sep} is used to detect sources on a weighted master detection image from all available ACS/WFC and WFC3/IR filters of a given field. We do not PSF match the \textit{HST} images; instead, we extract the photometry for each filter separately, with apertures of varying sizes, and then correct it to total flux densities by using curves of growth. We use a novel \textsc{golfir} algorithm, which relies on using the IRAC PSF - convolved high-resolution mosaics as a prior, to model and extract the IRAC photometry. Furthermore, we test the robustness of our derived photometry by comparing it to the publicly available HFF-DeepSpace \citep{shipley18} and ASTRODEEP \citep{astrodeep1,astrodeep2} catalogs in the Frontier Fields, and the photometric CLASH catalog of \citet{molino17}. In all cases we find results which are consistent, despite using a different approach. Moreover, compared to the aforementioned catalogs, the CHArGE data covers a $\sim \times 2$ area, and thus presents a substantial improvement in terms of the number of objects available.
\par
To derive photometric redshifts and stellar population parameters, we use the SED fitting software \textsc{EAZY} \citep{brammer08}. For the fields where uniform spectroscopic data is available (24/33), we achieve an average NMAD of 0.0406, for the photometric and spectroscopic redshifts across all fields, with an outlier fraction of 21 \%. To accompany our photometric redshifts, we provide stellar masses, star formation rates, extinctions, and other stellar population parameters based on the observed photometry. We also compute and use the rest-frame NUVrJ colors to separate our galaxies into potentially star-forming and quiescent samples. We find that our lensed cluster sample data contains an increased fraction of quiescent galaxies, compared to that of blind field observations. In addition to that, we use all the available magnification maps in the ALCS fields, and provide with our catalog magnification factors, where possible. We manually cross-match our data with the 145 S/N$>4$ ALMA detected galaxies, and provide upper limit measurement for a further $\sim$ 30,000 sources.
\par
These mosaics and catalogs, produced by the ALCS team, conclude one of the initial phases of the entire project, which, as outlined in \autoref{sec:sample}, will focus on multiple aspects and properties of faint sub-mm sources. These catalogues will also facilitate the detection, and further examination of the optically dark galaxy populations (e.g. \citealt{wang19,sun21,shu22}). 
As such these photometric catalogs can act as an important tool in designing future observations (e.g. with Keck/MOSFIRE, JWST, GMT) in an attempt to elucidate the key questions about the early onset of star formation, reionization and assembly of first galaxies.
\par
Both the \textit{HST}/\textit{Spitzer} mosaics, and photometric catalogs described in this work are publicly available in FITS format, through our repository \footnote{\url{https://github.com/dawn-cph/alcs-clusters}}. Alongside the photometric catalogs for each ALCS field, we include mosaics for all filters, detection images, segmentation maps, bright star masks, IRAC models and residuals. We also provide photometric redshifts and stellar population properties, as measured by \textsc{EAZY}, for each field. In the repository we also provide all the technical documentation regarding the source detection and modeling parameters, as well as notebooks to re-produce best fit \textsc{EAZY} SEDs.

\acknowledgments
We thank anonymous referee for a number of constructive suggestions which helped to improve this manuscript.
We thank Ian Smail for his helpful suggestions.
 The Cosmic Dawn Center is funded by the Danish National Research Foundation under grant No. 140. GEM acknowledges the Villum Fonden research grant 13160 “Gas to stars, stars to dust: tracing star formation across cosmic time,” grant 37440, “The Hidden Cosmos,” and the Cosmic Dawn Center of Excellence funded by the Danish National Research Foundation under the grant No. 140. FV acknowledges support from the Carlsberg Foundation research grant CF18-0388 “Galaxies: Rise And Death”. ID acknowledges support from the European Union’s Horizon 2020 research and innovation program under the Marie Sk\l{}odowska-Curie grant agreement No. 896225. ST and GB acknowledge support from the European Research Council (ERC) Consolidator   Grant funding scheme (project  ConTExt,  grant  No.   648179). 
 K. Kohno acknowledges the support by JSPS KAKENHI Grant Number JP17H06130 and the NAOJ ALMA Scientific Research Grant Number 2017-06B. FS acknowledges support from the NRAO Student Observing Support (SOS) award SOSPA7-022. DE acknowledges support from the Beatriz Galindo senior fellowship (BG20/00224) from the Spanish Ministry of Science and Innovation, projects PID2020-114414GB-100 and PID2020-113689GB-I00 financed by MCIN/AEI/10.13039/501100011033, project P20\_00334  financed by the Junta de Andaluc\'{i}a, and project A-FQM-510-UGR20 of the FEDER/Junta de Andaluc\'{i}a-Consejer\'{i}a de Transformaci\'{o}n Econ\'{o}mica, Industria, Conocimiento y Universidades. This work is based on observations taken by the RELICS Treasury Program (GO 14096) with the NASA/ESA HST, which is operated by the Association of Universities for Research in Astronomy, Inc., under NASA contract NAS5-26555.

\software{Astrodrizzle \citep{astrodrizzle}, EAZY \citep{brammer08}, FSPS \citep{conroy09}, GALFIT \citep{peng02,peng10}, grizli \citep{grizli}, golfir \citep{golfir}, MOPHONGO \citep{labbe13,labbe15}, sep \citep{sep}, SExtractor \citep{sextractor}}

\vspace{5mm}
\facilities{$HST$, $Spitzer$/IRAC, ALMA}
\clearpage
\appendix 

\section{Data Description} \label{sec:appendix_0}
Both the mosaics and photometric catalogs described in this work are publicly available in FITS format via our repository (\url{https://github.com/dawn-cph/alcs-clusters/}). The data are also available via the Electronic Research Data Archive\dataset[(ERDA)]{https://erda.ku.dk/archives/267ad68b71c7f89945180c8ba08f621d/published-archive.html} at the University of Copenhagen.
\par
For each field we make available the science - \texttt{sci} and inverse variance - \texttt{wht} mosaics for each \textit{HST} filter. The units of the filter mosaics are electrons/s, with the photometric calibration to CGS units provided in the \texttt{PHOTFLAM} ($f_\lambda$) and \texttt{PHOTFNU} ($f_\nu$) header keywords. For IRAC 3.6 $\mu$m and 4.5 $\mu$m we also provide the science \texttt{sci} and inverse variance - \texttt{wht} maps. In addition to that we include the model and the residual images for the \textsc{golfir} fit are included in the \texttt{model} files. The units for all IRAC images are given in $\mu$Jy.
\par
A listing of detection parameters from \textsc{sep} is
shown in \autoref{tab:det_par}. Each photometric catalog includes the source id, position, aperture and total photometry in all available filters, as well as photometric redshifts and physical parameters as computed by \textsc{EAZY}. The units of the photometry are in $\mu$Jy, which corresponds to the magnitude in the AB system of $23.9 - 2.5$ log10($f_\nu$ /$\mu$Jy).  In \autoref{tab:energy_corr} we provide the encircled energy corrections for all \textit{HST} filters within a $D=$0\farcs7 aperture. For each field we also compute the median $1\sigma$ depth from uncertainty on the total photometry. These values are presented in \autoref{tab:sigma}. All \textit{HST} data used in this work are listed in \autoref{tab:image_src}. IRAC data used in this work was collected during the following programs: HFF (Spitzer Frontier Fields; \citealt{spitzer_hff}), RELICS (Spitzer Reionization Lensing Cluster Survey; \citealt{spitzer_relics} ) and CLASH \citep{spitzer_clash}.

\begin{deluxetable}{cc}[h]
\tabcolsep=1.cm
\tablecaption{\label{tab:det_par}
\textsc{sep} Parameters Used for \textit{HST} Source Detection and Aperture Photometry.}
\tablehead{%
Parameter & Value}
\startdata
\hline
\texttt{BACK\_FILTERSIZE} & 4\farcs0 \\
\texttt{BACK\_FILTER} & 3 \\
\texttt{FILTER} & Y\\
\texttt{FILTER\_NAME} & F160W  \\
\texttt{CLEAN} & Y\\
\texttt{DEBLEND\_CONT} &  0.001\\
\texttt{DEBLEND\_NTHRESH} & 32 \\
\texttt{MINAREA} & 9 \\
\texttt{THRESHOLD} & 1.0 \\
\enddata
\end{deluxetable}

\begin{deluxetable}{cc}[h]
\tabcolsep=1.cm
\tablecaption{\label{tab:energy_corr}
Encircled Energy Correction for $D=$0.\farcs7 Apertures.}
\tablehead{%
Filter & Encircled Energy Fraction}
\startdata
\multicolumn{2}{c}{WFC3/NIR}\\
\hline
F105W & 0.839\\
F110W & 0.830\\
F125W & 0.825\\
F140W &	0.815\\
F160W &	0.803\\
\hline
\multicolumn{2}{c}{ACS/WFC}\\
\hline
F435W &	0.887\\
F475W &	0.893\\
F555W &	0.895\\
F606W &	0.896\\
F625W &	0.896\\
F775W &	0.894\\
F814W &	0.889\\
F850LP & 0.853\\
\hline
\multicolumn{2}{c}{WFC3/UVIS}\\
\hline
F275WU & 0.856\\
F336WU & 0.881\\
F390WU & 0.891\\
F438WU & 0.897\\
F606WU & 0.899\\
F625WU & 0.898\\
F814WU & 0.890\\
\enddata
\end{deluxetable}

\begin{deluxetable*}{cccccc}
\tablecaption{\label{tab:sigma}
	Effective Depths of \textit{HST}/\textit{Spitzer} Data}
\tablehead{%
Field	& \textit{HST}/f814w	& \textit{HST}/f125w	& \textit{HST}/f160w   & IRAC 3.6 $\mu$m   & IRAC 4.5 $\mu$m \\
& \multicolumn{5}{c}{Median $1\sigma$ depth $^\dagger$ [$\mu$Jy]}}
\startdata
\multicolumn{6}{c}{ALCS: Hubble Frontier Fields}\\
\hline \hline
Abell S1063 & 0.0097 & 0.0186 & 0.0210 & 0.0437 & 0.0332 \\
Abell 370 & 0.0075 & 0.0221 & 0.0270 & 0.0467 & 0.0408 \\
MACSJ0416.10-2403 & 0.0060 & 0.0183 & 0.0237 & 0.0417 & 0.0307 \\
Abell 2744 & 0.0087 & 0.0201 & 0.0256 & 0.051 & 0.0392 \\
MACSJ1149.5+2223 &  0.0037 & 0.0053 & 0.0078 & 0.0432 & 0.0358 \\
\hline
\multicolumn{6}{c}{ALCS: RELICS}\\
\hline \hline
RXCJ0032.1+1808  & 0.0210 & 0.0455 & 0.0292 & 0.1621 & 0.1390 \\
Abell 2537 & 0.0240 & 0.0496 & 0.0336 & 0.1725 & 0.1472 \\
Abell 3192 & 0.0129 & 0.0469 & 0.0277 & 0.1501 & 0.1093 \\
MACSJ0553.4-3342 &	0.0127 & 0.0439 & 0.0252 & 0.0963 & 0.0690 \\
RXC J0600.1-2007 &	0.0190 & 0.0477 & 0.0288 & 0.1581 & 0.1253 \\
RXC J0949.8+1707 &	0.0231 & 0.0576 & 0.0372 & 0.1631 & 0.1324 \\
MACSJ0257.1-2325 &	0.0107 & 0.0462 & 0.0287 & 0.1760 & 0.1628 \\
Abell 2163 & 0.0166 & 0.0496 & 0.0465 & 0.1265 & 0.1461 \\
PLCK G171.9-40.7 & 0.0263 & 0.0531 & 0.0351 & 0.1526 & 0.1354 \\
SMACSJ0723.3-7327 &	0.0219 & 0.0150 & 0.0173 & 0.0638 & 0.0483 \\
MACSJ0035.4-2015 &	0.0238 & 0.0397 & 0.0251 & 0.1634 & 0.1292 \\
MACSJ0417.5-1154 & 0.0228 & 0.0448 & 0.0371 & 0.1689 & 0.1263 \\
MACSJ0159.8-0849 &	0.0218 & 0.0411 & 0.0257 & 0.1646 & 0.1405 \\
ACT-CLJ0102-49151 &	0.0189 & 0.0400 & 0.0234 & 0.0688 & 0.0533 \\
AbellS295 	& 0.0160 & 0.0426 & 0.0250 & 0.0839 & 0.0631 \\
RXC J2211.7-0350 & 0.0237 & 0.0527 & 0.0340 & 0.1495 & 0.1258\\
\hline
\multicolumn{6}{c}{ALCS: CLASH}\\
\hline \hline
Abell 383 & 0.0147 & 0.0211 & 0.0211 & 0.1278 & 0.1647 \\
MACS1206.2-0847 & 0.0154 & 0.0203 & 0.0225 & 0.1872 & 0.1495 \\
MACS1423.8+2404 & 0.0107 & 0.0132 & 0.0144 & 0.0632 & 0.0516 \\
MACS1931.8-2635 & 0.0195 & 0.0221 & 0.0220 & 1.2199 & 0.1568 \\
RXJ 1347-1145 & 0.0115 & 0.0201 & 0.0152 & 0.0529 & 0.0482 \\
MACS1311.0-0310 & 0.0161 & 0.0201 & 0.0211 & 0.1874 & 0.1609 \\
MACS1115.9+0129 & 0.0165 & 0.0204 & 0.0214 & 0.1024 & 0.0629 \\
MACS0429.6-0253 & 0.0147 & 0.0181 & 0.0206 & 0.1786 & 0.1496 \\
RXJ2129.7+0005 & 0.0151 & 0.0170 & 0.0174 & 0.1608 & 0.1299 \\ 
MACS0329.7-0211 & 0.0150 & 0.0183 & 0.0191 & 0.1790 & 0.1493 \\
MACS2129.4-0741 & 0.0113 & 0.0231 & 0.0209 & 0.0616 & 0.0533 \\
Abell 209 & 0.0154 & 0.0203 & 0.0183 & 0.1223 & 0.1384 \\
\enddata
\begin{tablenotes}
$\dagger$ \footnotesize{Computed directly from the final photometric catalogs.}
\end{tablenotes}
\end{deluxetable*}

\clearpage

\startlongtable
\begin{deluxetable*}{ccccc}
\tablecaption{\label{tab:image_src}
	\textit{HST} Image Sources}
\tablehead{%
Field & Filters & Instrument & Proposal ID & Proposal PI}
\startdata
\multicolumn{5}{c}{ALCS: Hubble Frontier Fields}\\
\hline \hline
Abell S1063 & F435W, F475W, F606W, F625W &  ACS/WFC & 12458 & Postman, Marc \\
& F775W, F814W, F850LP &  ACS/WFC & 12458 & Postman, Marc \\
& F814W &  ACS/WFC & 13063 & Riess, Adam \\
& F814W  &  ACS/WFC & 13459 & Treu, Tommaso L. \\
& F435W, F606W, F814W &  ACS/WFC & 14037 & Lotz, Jennifer \\
& F435W, F606W &  ACS/WFC & 14209 & Siana, Brian \\
& F606W, F814W &  ACS/WFC & 15117 & Steinhardt, Charles L. \\
& F606W, F814W &  ACS/WFC & 15936 & Kelly, Patrick \\
& F475W &  ACS/WFC & 15940 & Ribeiro, Bruno \\
& F606W, F814W &  ACS/WFC & 16278 & Kelly, Patrick \\
& F105W, F110W, F125W, F140W, F160W &  WFC3/IR & 12458 & Postman, Marc \\
& F125W, F160W &  WFC3/IR & 13063 & Riess, Adam \\
& F105W, F140W &  WFC3/IR & 13459 & Treu, Tommaso L. \\
& F105W, F125W, F140W, F160W &  WFC3/IR & 14037 & Lotz, Jennifer \\
& F105W, F125W, F160W &  WFC3/IR & 15117 & Steinhardt, Charles L. \\
& F110W &  WFC3/IR & 16729 & Kelly, Patrick \\
& F225W, F275W, F336W, F390W &  WFC3/UVIS & 12458 & Postman, Marc \\
& F275W, F336W &  WFC3/UVIS & 14209 & Siana, Brian \\
& F225W &  WFC3/UVIS & 15940 & Ribeiro, Bruno \\
& F225W, F275W &  WFC3/UVIS & 16239 & Foley, Ryan \\
& F606W &  WFC3/UVIS & 16729 & Kelly, Patrick \\
\hline
Abell 370 & F475W, F625W, F814W &  ACS/WFC & 11507 & Noll, Keith S. \\
& F475W &  ACS/WFC & 11582 & Blain, Andrew \\
& F814W &  ACS/WFC & 11591 & Kneib, Jean-Paul Richard \\
& F814W  &  ACS/WFC & 13459 & Treu, Tommaso L. \\
& F814W &  ACS/WFC & 13790 & Rodney, Steve \\
& F435W, F606W, F814W &  ACS/WFC & 14038 & Lotz, Jennifer \\
& F435W, F606W &  ACS/WFC & 14209 & Siana, Brian \\
& F606W, F814W &  ACS/WFC & 15117 & Steinhardt, Charles L. \\
& F475W &  ACS/WFC & 15940 & Ribeiro, Bruno \\
& F606W, F814W &  ACS/WFC & 16278 & Kelly, Patrick \\
& F140W &  WFC3/IR & 11108 & Hu, Esther M. \\
& F110W, F160W &  WFC3/IR & 11591 & Kneib, Jean-Paul Richard \\
& F160W &  WFC3/IR & 12880 & Riess, Adam \\
& F105W, F140W &  WFC3/IR & 13459 & Treu, Tommaso L. \\
& F110W, F125W, F160W &  WFC3/IR & 13790 & Rodney, Steve \\
& F105W, F125W, F140W, F160W &  WFC3/IR & 14038 & Lotz, Jennifer \\
& F160W &  WFC3/IR & 14216 & Kirshner, Robert P. \\
& F105W, F125W, F160W &  WFC3/IR & 15117 & Steinhardt, Charles L. \\
& F555W, F814W &  WFC3/UVIS & 12880 & Riess, Adam \\
& F625W, F814W &  WFC3/UVIS & 13790 & Rodney, Steve \\
& F275W, F336W &  WFC3/UVIS & 14209 & Siana, Brian \\
& F225W &  WFC3/UVIS & 15940 & Ribeiro, Bruno \\
& F275W &  WFC3/UVIS & 16741 & Galbany, Lluis  \\
\hline
MACSJ0416.10-2403 & F435W, F475W, F606W, F625W &  ACS/WFC & 12459 & Postman, Marc \\
& F775W, F814W, F850LP &  ACS/WFC & 12459 & Postman, Marc \\
& F606W, F814W &  ACS/WFC & 13386 & Rodney, Steve \\
& F814W &  ACS/WFC & 13459 & Treu, Tommaso L. \\
& F435W, F606W, F814W &  ACS/WFC & 13496 & Lotz, Jennifer \\
& F435W, F606W &  ACS/WFC & 14209 & Siana, Brian \\
& F606W, F814W &  ACS/WFC & 15117 & Steinhardt, Charles L. \\
& F606W, F814W &  ACS/WFC & 15936 & Kelly, Patrick \\
& F475W &  ACS/WFC & 15940 & Ribeiro, Bruno \\
& F606W, F814W &  ACS/WFC & 16278 & Kelly, Patrick \\
& F105W, F110W, F125W, F140W, F160W &  WFC3/IR & 12459 & Postman, Marc \\
& F105W, F125W, F160W &  WFC3/IR & 13386 & Rodney, Steve \\
& F105W, F140W  &  WFC3/IR & 13459 & Treu, Tommaso L. \\
& F105W, F125W, F140W, F160W &  WFC3/IR & 13496 & Lotz, Jennifer \\
& F105W, F125W, F160W &  WFC3/IR & 15117 & Steinhardt, Charles L. \\
& F105W &  WFC3/IR & 16729 & Kelly, Patrick \\
& F225W, F275W, F336W, F390W &  WFC3/UVIS & 12459 & Postman, Marc \\
& F275W, F336W &  WFC3/UVIS & 14209 & Siana, Brian \\
& F225W &  WFC3/UVIS & 15940 & Ribeiro, Bruno \\
& F606W &  WFC3/UVIS & 16729 & Kelly, Patrick \\
\hline
Abell 2744 & F435W, F606W, F814W &  ACS/WFC & 11689 & Dupke, Renato A. \\
& F606W, F814W  &  ACS/WFC & 13386 & Rodney, Steve \\
& F435W, F606W &  ACS/WFC & 13389 & Siana, Brian \\
& F814W  &  ACS/WFC & 13459 & Treu, Tommaso L. \\
& F435W, F606W, F814W &  ACS/WFC & 13495 & Lotz, Jennifer \\
& F606W, F814W &  ACS/WFC & 15117 & Steinhardt, Charles L. \\
& F475W &  ACS/WFC & 15940 & Ribeiro, Bruno \\
& F606W, F775W &  ACS/WFC & 17231 & Treu, Tommaso L. \\
& F105W, F125W, F160W &  WFC3/IR & 13386 & Rodney, Steve \\
& F105W, F140W &  WFC3/IR & 13459 & Treu, Tommaso L. \\
& F105W, F125W, F140W, F160W &  WFC3/IR & 13495 & Lotz, Jennifer \\
& F105W, F125W, F160W &  WFC3/IR & 15117 & Steinhardt, Charles L. \\
& F275W, F336W &  WFC3/UVIS & 13389 & Siana, Brian \\
& F225W &  WFC3/UVIS & 15940 & Ribeiro, Bruno \\
\hline
MACSJ1149.5+2223 & F814W &  ACS/WFC & 10493 & Gal-Yam, Avishay \\
& F435W, F475W, F606W, F625W &  ACS/WFC & 12068 & Postman, Marc \\
& F775W, F850LP &  ACS/WFC & 12068 & Postman, Marc \\
& F435W, F606W &  ACS/WFC & 13389 & Siana, Brian \\
& F814W &  ACS/WFC & 13459 & Treu, Tommaso L. \\
& F435W, F606W, F814W &  ACS/WFC & 13504 & Lotz, Jennifer \\
& F606W, F814W &  ACS/WFC & 13790 & Rodney, Steve \\
& F606W, F814W &  ACS/WFC & 14199 & Kelly, Patrick \\
& F435W, F475W, F606W, F814W &  ACS/WFC & 14208 & Rodney, Steve \\
& F606W, F814W &  ACS/WFC & 14872 & Kelly, Patrick \\
& F606W, F814W &  ACS/WFC & 15117 & Steinhardt, Charles Louis \\
& F606W, F814W &  ACS/WFC & 15308 & Gonzalez, Anthony Hernan \\
& F606W, F814W &  ACS/WFC & 15936 & Kelly, Patrick \\
& F475W &  ACS/WFC & 15940 & Ribeiro, Bruno \\
& F775W &  ACS/WFC & 9480 & Rhodes, Jason D. \\
& F775W &  ACS/WFC & 9584 & Sparks, William B. \\
& F555W, F814W &  ACS/WFC & 9722 & Ebeling, Harald \\
& F105W, F110W, F125W, F140W, F160W &  WFC3/IR & 12068 & Postman, Marc \\
& F105W, F140W &  WFC3/IR & 13459 & Treu, Tommaso L. \\
& F105W, F125W, F140W, F160W &  WFC3/IR & 13504 & Lotz, Jennifer \\
& F105W, F125W, F140W, F160W &  WFC3/IR & 13767 & Trenti, Michele \\
& F105W, F125W, F160W &  WFC3/IR & 13790 & Rodney, Steve \\
& F125W, F160W &  WFC3/IR & 14041 & Kelly, Patrick \\
& F105W, F125W, F160W &  WFC3/IR & 14199 & Kelly, Patrick \\
& F105W, F125W, F160W &  WFC3/IR & 14208 & Rodney, Steve \\
& F125W, F160W &  WFC3/IR & 14528 & Kelly, Patrick \\
& F110W, F125W, F160W &  WFC3/IR & 14872 & Kelly, Patrick \\
& F110W &  WFC3/IR & 14922 & Kelly, Patrick \\
& F105W, F125W, F160W &  WFC3/IR & 15117 & Steinhardt, Charles L.\\
& F105W, F160W &  WFC3/IR & 15308 & Gonzalez, Anthony Hernan \\
& F225W, F275W, F336W, F390W &  WFC3/UVIS & 12068 & Postman, Marc \\
& F275W, F336W &  WFC3/UVIS & 13389 & Siana, Brian \\
& F814W &  WFC3/UVIS & 13790 & Rodney, Steve \\
& F606W, F814W &  WFC3/UVIS & 14041 & Kelly, Patrick \\
& F275W, F336W, F606W &  WFC3/UVIS & 14199 & Kelly, Patrick \\
& F336W &  WFC3/UVIS & 14208 & Rodney, Steve \\
& F606W &  WFC3/UVIS & 14528 & Kelly, Patrick \\
& F606W &  WFC3/UVIS & 14872 & Kelly, Patrick \\
& F606W &  WFC3/UVIS & 14922 & Kelly, Patrick \\
& F225W &  WFC3/UVIS & 15940 & Ribeiro, Bruno \\
\\
\\
\\
\hline
\multicolumn{5}{c}{ALCS: CLASH}\\
\hline \hline
Abell 383 & F435W, F475W, F606W &  ACS/WFC & 12065 & Postman, Marc \\
&  F625W, F775W, F814W, F850LP &  ACS/WFC & 12065 & Postman, Marc \\
& F606W, F850LP &  ACS/WFC & 12099 & Riess, Adam \\
& F105W, F110W, F125W, F140W, F160W &  WFC3/IR & 12065 & Postman, Marc \\
& F105W, F160W &  WFC3/IR & 12099 & Riess, Adam \\
& F105W, F125W, F160W &  WFC3/IR & 12360 & Perlmutter, Saul \\
& F225W, F275W, F336W &  WFC3/UVIS & 12065 & Postman, Marc \\
& F390W, F814W &  WFC3/UVIS & 12065 & Postman, Marc \\
& F814W &  WFC3/UVIS & 12099 & Riess, Adam \\
\hline
MACS1206.2-0847 & F606W &  ACS/WFC & 10491 & Ebeling, Harald \\
& F435W, F475W, F606W, F625W &  ACS/WFC & 12069 & Postman, Marc \\
& F775W, F814W, F850LP &  ACS/WFC & 12069 & Postman, Marc \\
& F105W, F110W, F125W, F140W, F160W &  WFC3/IR & 12069 & Postman, Marc \\
& F110W &  WFC3/IR & 16729 & Kelly, Patrick \\
& F225W, F275W, F336W, F390W &  WFC3/UVIS & 12069 & Postman, Marc \\
& F336W &  WFC3/UVIS & 15271 & Ferguson, Harry C. \\
& F606W &  WFC3/UVIS & 16729 & Kelly, Patrick \\
\hline
MACS1423.8+2404 & F814W &  ACS/WFC & 10493 & Gal-Yam, Avishay \\
& F435W, F475W, F606W, F775W, F850LP &  ACS/WFC & 12790 & Postman, Marc \\
& F850LP &  ACS/WFC & 13063 & Riess, Adam \\
& F606W, F775W, F814W, F850LP &  ACS/WFC & 13386 & Rodney, Steve \\
& F814W &  ACS/WFC & 13459 & Treu, Tommaso L. \\
& F814W &  ACS/WFC & 15444 & Barth, Aaron J. \\
& F435W, F606W &  ACS/WFC & 16667 & Bradac, Marusa \\
& F555W, F814W &  ACS/WFC & 9722 & Ebeling, Harald \\
& F105W, F110W, F125W, F140W, F160W &  WFC3/IR & 12790 & Postman, Marc \\
& F125W, F160W &  WFC3/IR & 13063 & Riess, Adam \\
& F105W, F125W, F140W, F160W &  WFC3/IR & 13386 & Rodney, Steve \\
& F105W, F140W &  WFC3/IR & 13459 & Treu, Tommaso L. \\
& F225W, F275W, F336W, F390W &  WFC3/UVIS & 12790 & Postman, Marc \\
& F606W &  WFC3/UVIS & 13386 & Rodney, Steve \\
& F438W, F606W &  WFC3/UVIS & 16667 & Bradac, Marusa \\
\hline
MACS1931.8-2635 & F435W, F475W, F606W, F625W &  ACS/WFC & 12456 & Postman, Marc \\
& F775W, F814W, F850LP &  ACS/WFC & 12456 & Postman, Marc \\
& F105W, F110W, F125W, F140W, F160W &  WFC3/IR & 12456 & Postman, Marc \\
& F225W, F275W, F336W, F390W &  WFC3/UVIS & 12456 & Postman, Marc  \\
\hline
RXJ 1347-1145 & F475W, F814W, F850LP &  ACS/WFC & 10492 & Erben, Thomas \\
& F814W &  ACS/WFC & 11591 & Kneib, Jean-Paul Richard \\
& F435W, F606W, F625W, F775W, F850LP &  ACS/WFC & 12104 & Postman, Marc \\
& F814W &  ACS/WFC & 13459 & Treu, Tommaso L. \\
& F110W, F160W &  WFC3/IR & 11591 & Kneib, Jean-Paul Richard \\
& F105W, F110W, F125W, F140W, F160W &  WFC3/IR & 12104 & Postman, Marc \\
& F105W, F160W &  WFC3/IR & 13386 & Rodney, Steve \\
& F105W, F140W &  WFC3/IR & 13459 & Treu, Tommaso L. \\
& F110W &  WFC3/IR & 16729 & Kelly, Patrick \\
& F225W, F275W, F336W, F390W &  WFC3/UVIS & 12104 & Postman, Marc \\
& F606W &  WFC3/UVIS & 13386 & Rodney, Steve \\
& F606W &  WFC3/UVIS & 16729 & Kelly, Patrick \\
\hline
MACS1311.0-0310 & F435W, F475W, F606W, F625W &  ACS/WFC & 12789 & Postman, Marc \\
& F775W, F814W, F850LP &  ACS/WFC & 12789 & Postman, Marc \\
& F105W, F110W, F125W, F140W, F160W &  WFC3/IR & 12789 & Postman, Marc \\
& F225W, F275W, F336W, F390W &  WFC3/UVIS & 12789 & Postman, Marc \\
\hline
MACS1115.9+0129 & F606W &  ACS/WFC & 10491 & Ebeling, Harald \\
& F435W, F475W, F606W, F625W &  ACS/WFC & 12453 & Postman, Marc \\
& F775W, F814W, F850LP &  ACS/WFC & 12453 & Postman, Marc \\
& F105W, F110W, F125W, F140W, F160W &  WFC3/IR & 12453 & Postman, Marc \\
& F110W &  WFC3/IR & 16729 & Kelly, Patrick \\
& F225W, F275W, F336W, F390W &  WFC3/UVIS & 12453 & Postman, Marc \\
& F606W &  WFC3/UVIS & 16729 & Kelly, Patrick \\
\hline
MACS0429.6-0253 & F435W, F475W, F606W, F625W &  ACS/WFC & 12788 & Postman, Marc \\
& F775W, F814W, F850LP &  ACS/WFC & 12788 & Postman, Marc \\
& F105W, F110W, F125W, F140W, F160W &  WFC3/IR & 12788 & Postman, Marc \\
& F225W, F275W, F336W, F390W &  WFC3/UVIS & 12788 & Postman, Marc \\
& F225W, F336W &  WFC3/UVIS & 16173 & Tremblay, Grant R. \\
\hline
RXJ2129.7+0005 & F606W &  ACS/WFC & 10588 & Brotherton, Michael S. \\
& F435W, F475W, F606W, F625W &  ACS/WFC & 12457 & Postman, Marc \\
& F775W, F814W, F850LP &  ACS/WFC & 12457 & Postman, Marc \\
& F555W, F775W, F850LP &  ACS/WFC & 12461 & Riess, Adam \\
& F105W, F110W, F125W, F140W, F160W &  WFC3/IR & 12457 & Postman, Marc \\
& F105W, F125W, F160W &  WFC3/IR & 12461 & Riess, Adam \\
& F225W, F275W, F336W, F390W &  WFC3/UVIS & 12457 & Postman, Marc \\
\hline
MACS0329.7-0211 & F435W, F475W, F606W, F625W &  ACS/WFC & 12452 & Postman, Marc \\
& F775W, F814W, F850LP &  ACS/WFC & 12452 & Postman, Marc \\
& F105W, F110W, F125W, F140W, F160W &  WFC3/IR & 12452 & Postman, Marc \\
& F225W, F275W, F336W, F390W &  WFC3/UVIS & 12452 & Postman, Marc \\
\hline
MACS2129.4-074 & F814W &  ACS/WFC & 10493 & Gal-Yam, Avishay \\
& F775W, F814W, F850LP &  ACS/WFC & 12099 & Riess, Adam \\
& F435W, F475W, F606W, F625W, F775W &  ACS/WFC & 12100 & Postman, Marc \\
& F850LP &  ACS/WFC & 12100 & Postman, Marc \\
& F606W, F814W &  ACS/WFC & 13386 & Rodney, Steve \\
& F814W &  ACS/WFC & 13459 & Treu, Tommaso L. \\
& F814W &  ACS/WFC & 13790 & Rodney, Steve \\
& F555W, F814W &  ACS/WFC & 9722 & Ebeling, Harald \\
& F125W, F160W &  WFC3/IR & 12099 & Riess, Adam \\
& F105W, F110W, F125W, F140W, F160W &  WFC3/IR & 12100 & Postman, Marc \\
& F105W, F125W, F140W, F160W &  WFC3/IR & 13386 & Rodney, Steve \\
& F105W, F140W &  WFC3/IR & 13459 & Treu, Tommaso L. \\
& F125W, F160W &  WFC3/IR & 13790 & Rodney, Steve \\
& F140W &  WFC3/IR & 15663 & Akhshik, Mohammad \\
& F225W, F275W, F336W &  WFC3/UVIS & 12099 & Riess, Adam \\
& F225W, F275W, F336W, F390W &  WFC3/UVIS & 12100 & Postman, Marc \\
\hline
Abell 209 & F435W, F475W, F606W, F625W &  ACS/WFC & 12451 & Postman, Marc \\
& F775W, F814W, F850LP &  ACS/WFC & 12451 & Postman, Marc \\
& F105W, F110W, F125W, F140W, F160W &  WFC3/IR & 12451 & Postman, Marc \\
& F225W, F275W, F336W, F390W &  WFC3/UVIS & 12451 & Postman, Marc \\
\\
\hline
\multicolumn{5}{c}{ALCS: RELICS}\\
\hline \hline
RXCJ0032.1+1808 & F606W, F814W & ACS/WFC & 12166 & Ebeling, Harald \\
& F435W, F606W, F814W & ACS/WFC & 14096 & Coe, Dan \\
& F105W, F125W, F140W, F160W& WFC3/IR & 14096 & Coe, Dan \\
\hline
Abell 2537 & F435W, F814W& ACS/WFC & 14096 & Coe, Dan\\
& F606W &  ACS/WFC & 9270 & Allen, Steven W.\\
& F775W &  ACS/WFC & 9575 & Sparks, William B. \\
& F775W &  ACS/WFC & 9984 & Rhodes, Jason D. \\
& F105W, F125W, F140W, F160W &  WFC3/IR & 14096 & Coe, Dan\\
\hline
Abell 3192 & F606W &  ACS/WFC & 10881 & Smith, Graham \\
& F435W, F606W, F814W &  ACS/WFC & 12313 & Ebeling, Harald \\
& F105W, F125W, F140W, F160W &  WFC3/IR & 14096 & Coe, Dan \\
\hline
MACSJ0553.4-3342  & F435W, F606W, F814W &  ACS/WFC & 12362 & Ebeling, Harald \\
& F105W, F125W, F140W, F160W &  WFC3/IR & 14096 & Coe, Dan \\
\hline
RXC J0600.1-2007 & F814W &  ACS/WFC & 12884 & Ebeling, Harald \\
& F435W, F606W, F814W &  ACS/WFC & 14096 & Coe, Dan \\
& F105W, F125W, F140W, F160W &  WFC3/IR & 14096 & Coe, Dan \\
\hline
RXC J0949.8+1707 & F606W &  ACS/WFC & 10491 & Ebeling, Harald \\
& F814W &  ACS/WFC & 12166 & Ebeling, Harald \\
& F435W, F606W, F814W &  ACS/WFC & 14096 & Coe, Dan \\
& F110W &  WFC3/IR & 14047 & Reines, Amy E. \\
& F105W, F125W, F140W, F160W &  WFC3/IR & 14096 & Coe, Dan \\
\hline
MACSJ0257.1-2325 & F814W &  ACS/WFC & 10493 & Gal-Yam, Avishay \\
& F814W &  ACS/WFC & 10793 & Gal-Yam, Avishay \\
& F435W &  ACS/WFC & 14096 & Coe, Dan \\
& F555W, F814W &  ACS/WFC & 9722 & Ebeling, Harald \\
& F105W, F125W, F140W, F160W &  WFC3/IR & 14096 & Coe, Dan \\
\hline
Abell 2163 & F435W, F606W, F814W &  ACS/WFC & 12253 & Clowe, Douglas \\
& F105W, F125W, F140W, F160W &  WFC3/IR & 14096 & Coe, Dan \\
\hline
PLCK G171.9-40.7 & F435W, F606W, F814W &  ACS/WFC & 14096 & Coe, Dan \\ 
& F105W, F125W, F140W, F160W &  WFC3/IR & 14096 & Coe, Dan \\
\hline
SMACSJ0723.3-7327 & F606W &  ACS/WFC & 12166 & Ebeling, Harald \\
& F814W &  ACS/WFC & 12884 & Ebeling, Harald \\
& F435W, F606W, F814W &  ACS/WFC & 14096 & Coe, Dan \\
& F105W, F125W, F140W, F160W &  WFC3/IR & 14096 & Coe, Dan \\
\hline
MACSJ0035.4-2015 & F606W &  ACS/WFC & 10491 & Ebeling, Harald \\
& F814W &  ACS/WFC & 12884 & Ebeling, Harald \\ 
& F435W, F606W, F814W &  ACS/WFC & 14096 & Coe, Dan \\
& F105W, F125W, F140W, F160W &  WFC3/IR & 14096 & Coe, Dan \\
\hline
MACSJ0417.5-1154 & F814W &  ACS/WFC & 12009 & von der Linden, Anja \\
& F435W &  ACS/WFC & 14096 & Coe, Dan \\
& F435W, F606W &  ACS/WFC & 16667 & Bradac, Marusa \\
& F105W, F125W, F140W, F160W &  WFC3/IR & 14096 & Coe, Dan \\
& F606W &  WFC3/UVIS & 12009 & von der Linden, Anja \\
& F438W, F606W, F814W &  WFC3/UVIS & 16667 & Bradac, Marusa \\
& F606W &  WFC3/UVIS & 16863 & Anderson, Jay \\
\hline
MACSJ0159.8-0849 & F606W &  ACS/WFC & 12166 & Ebeling, Harald \\
& F435W, F606W, F814W &  ACS/WFC & 14096 & Coe, Dan \\
& F105W, F125W, F140W, F160W &  WFC3/IR & 14096 & Coe, Dan \\
\hline 
ACT-CLJ0102-49151 & F606W, F814W &  ACS/WFC & 12477 & High, Fredrick W. \\
& F625W, F775W, F850LP &  ACS/WFC & 12755 & Hughes, John P. \\
& F435W &  ACS/WFC & 14096 & Coe, Dan \\
& F606W &  ACS/WFC & 14153 & Hughes, John P. \\
& F105W, F125W, F140W, F160W &  WFC3/IR & 14096 & Coe, Dan \\
& F140W &  WFC3/IR & 16773 & Glazebrook, Karl \\
\hline
AbellS295 & F435W, F606W, F814W &  ACS/WFC & 13514 & Pacaud, Florian \\
& F105W, F125W, F140W, F160W &  WFC3/IR & 14096 & Coe, Dan \\
& F105W &  WFC3/IR & 16729 & Kelly, Patrick \\
& F606W &  WFC3/UVIS & 16729 & Kelly, Patrick \\
\hline
RXC J2211.7-0350 & F606W &  ACS/WFC & 12166 & Ebeling, Harald \\ & F435W, F606W, F814W &  ACS/WFC & 14096 & Coe, Dan \\
& F105W, F125W, F140W, F160W &  WFC3/IR & 14096 & Coe, Dan \\
& F475W &  WFC3/UVIS & 11565 & Lepine, Sebastien \\
\hline
\enddata
\begin{tablenotes}
\end{tablenotes}
\end{deluxetable*}

\clearpage

\section{Quality and Consistency Verification} \label{sec:appendix_1}
In this section we present the comparison between \textit{HST} broad-band photometry measurements, and low-resolution IRAC model photometry to the publicly available catalogs covering the ALCS fields. These include the HFF-DeepSpace and ASTRODEEP catalogs in the Hubble Frontier Fields \citep{astrodeep1,astrodeep2,shipley18}, which most closely match our photometric baseline and also process the IRAC photometry. In addition we also carry out the comparison between our photometry and the \citealt{molino17} catalogs, in the CLASH fields, albeit only for \textit{HST} data. The comparison is carried out on a per-filter basis, and also includes the difference between the derived colors. Where appropriate, we also contrast the area weighted number counts in the detection bands.

\begin{figure*}
\begin{center}
\includegraphics[width=.98\textwidth]{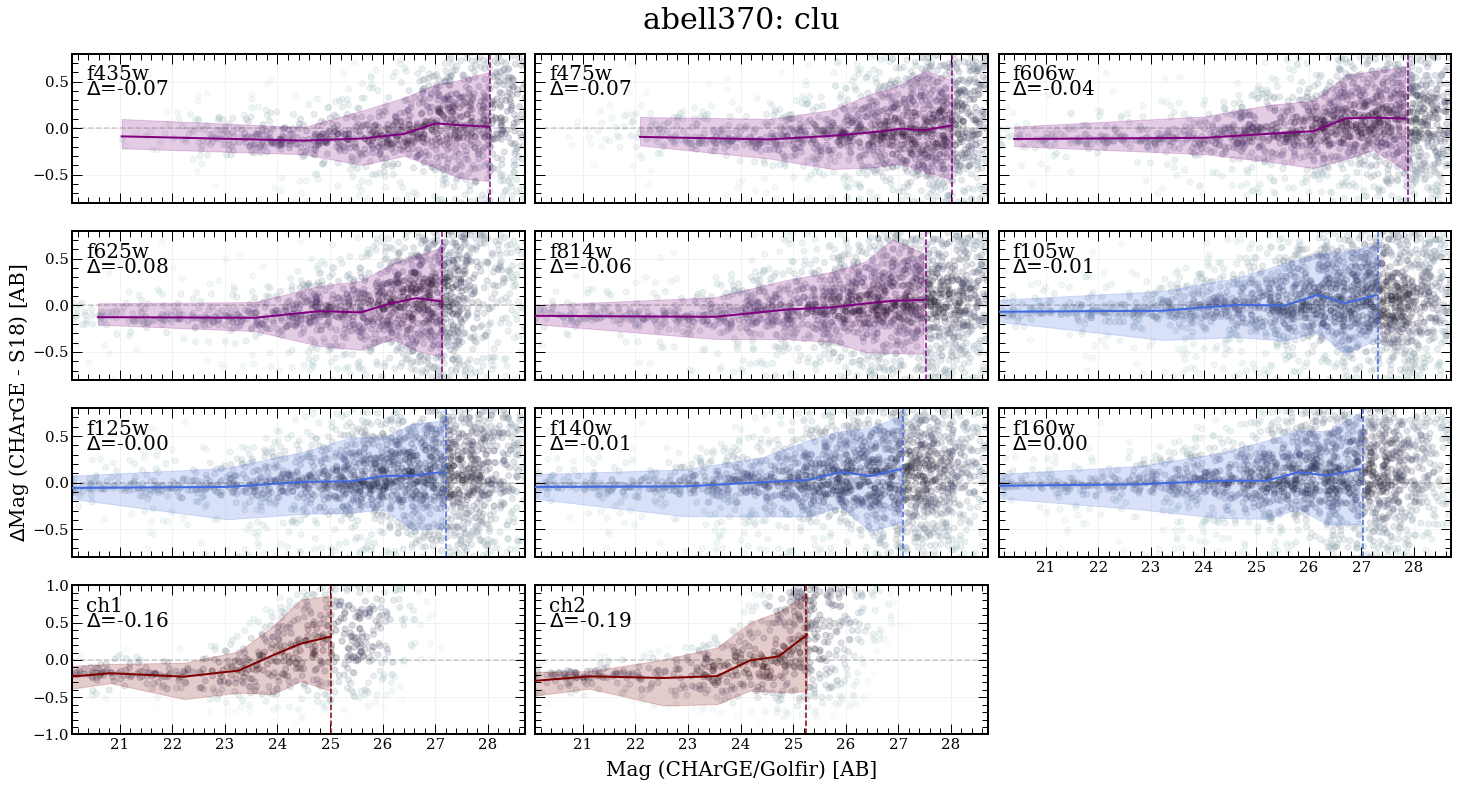}
\includegraphics[width=.98\textwidth]{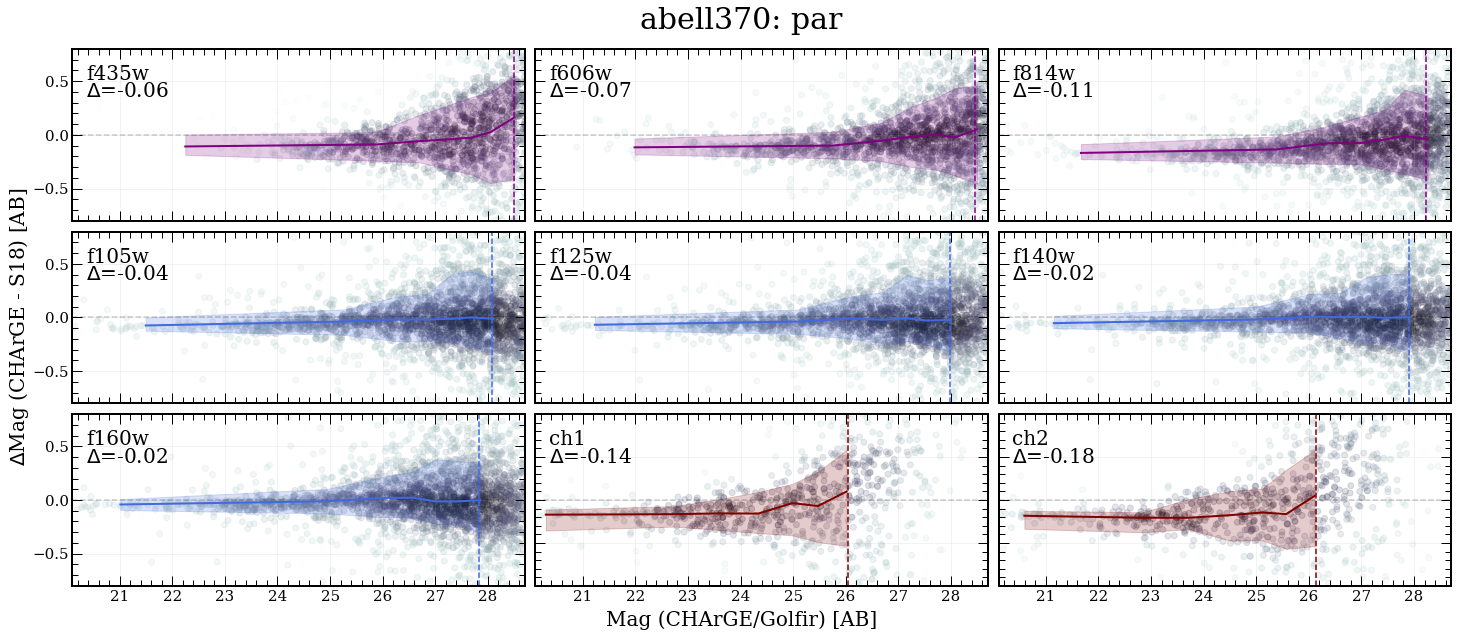}
\caption{The difference between broad-band magnitude measured in our catalog and \citealt{shipley18},
for the cluster (top) and parallel (bottom) parts of Abell 370 field. We only compare the objects in the central part of the field. For more clarity the colors of shaded regions correspond to $HST$/ACS (purple), $HST$/WFC3 (blue), and $Spitzer$/IRAC (maroon). The difference in magnitude $\Delta$Mag is shown by scattered circles, which are colored based on the density of sources around them. The overlaid solid lines correspond to the binned median, that are selected to contain the same amount of objects in each bin. The shaded envelope captures the 68\% of points per magnitude bin. The vertical dashed lines
correspond to the 1$\sigma$ depth limit for each band. The median $\Delta$Mag for galaxies brighter than the depth limit is shown on each panel.}
\label{fig:deltamag}
\end{center}
\end{figure*}

\begin{figure*}
\begin{center}
\includegraphics[width=.9\textwidth]{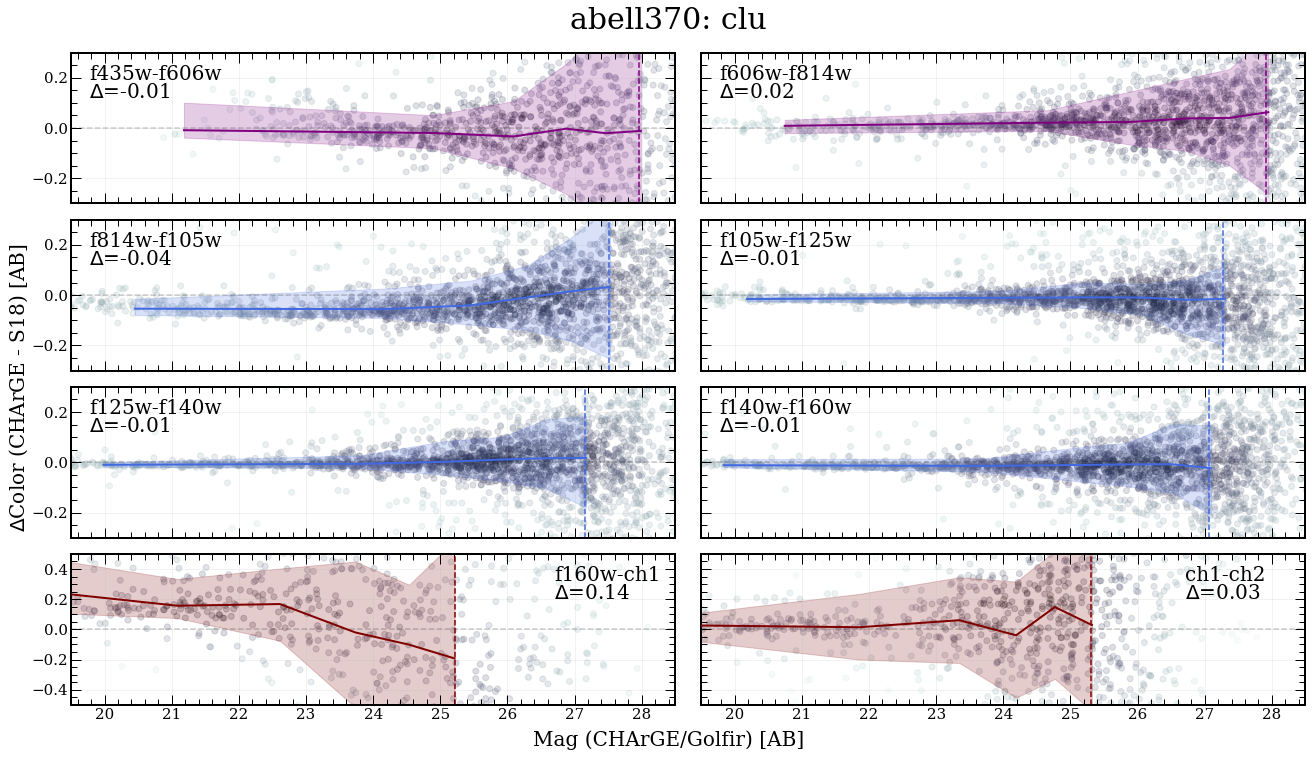}
\includegraphics[width=.9\textwidth]{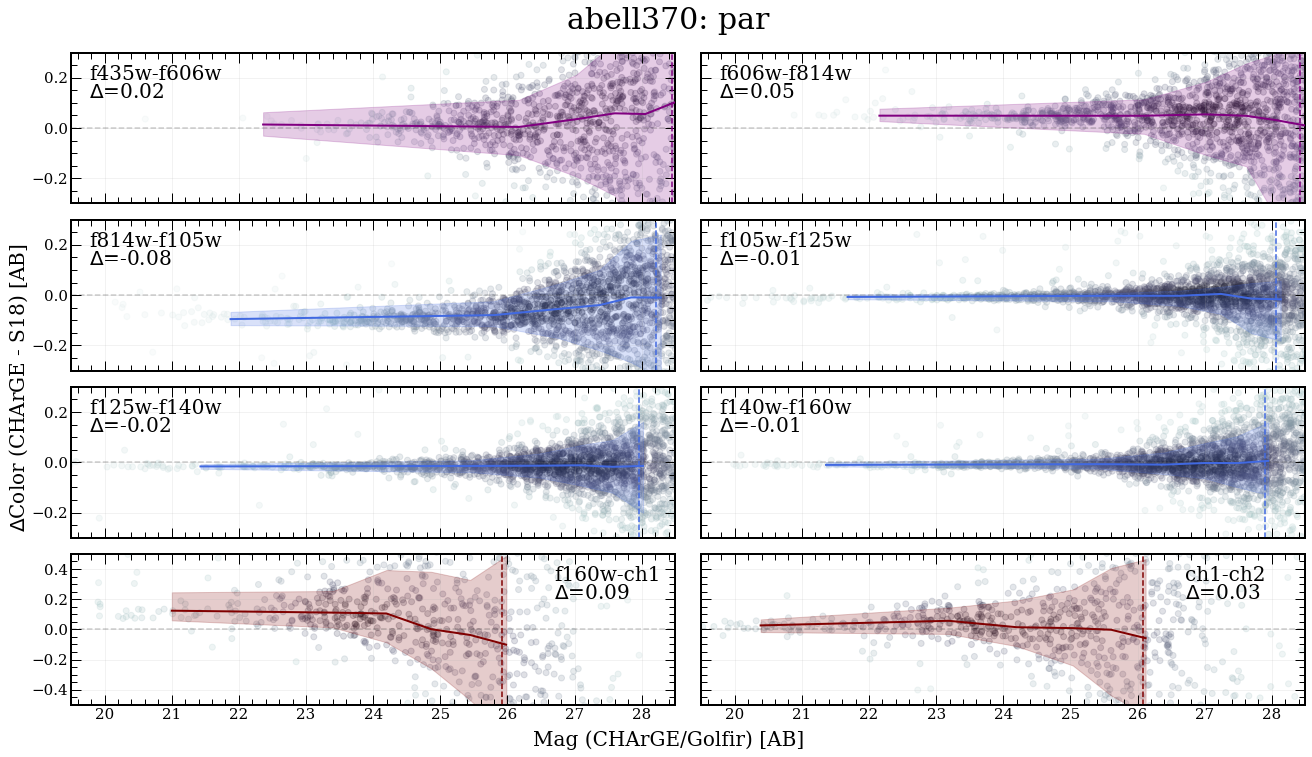}
\caption{The difference between broad-band color computed in our catalog and \citealt{shipley18}, for the cluster (top) and parallel (bottom) parts of Abell 370 field. The colors and symbols are the same as those in \autoref{fig:deltamag}.}
\label{fig:delta_color}
\end{center}
\end{figure*}

\begin{figure*}
\begin{center}
\includegraphics[width=.95\textwidth]{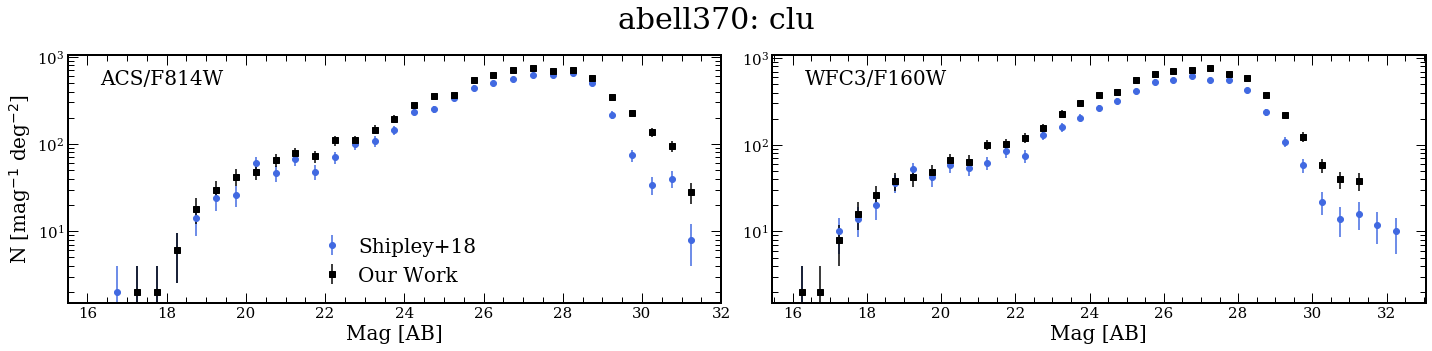}
\includegraphics[width=.95\textwidth]{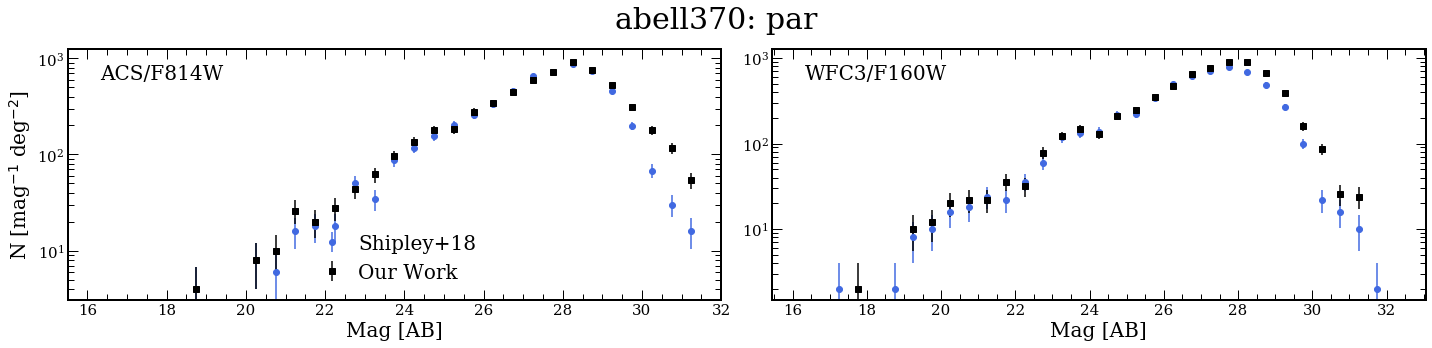}
\includegraphics[width=.95\textwidth]{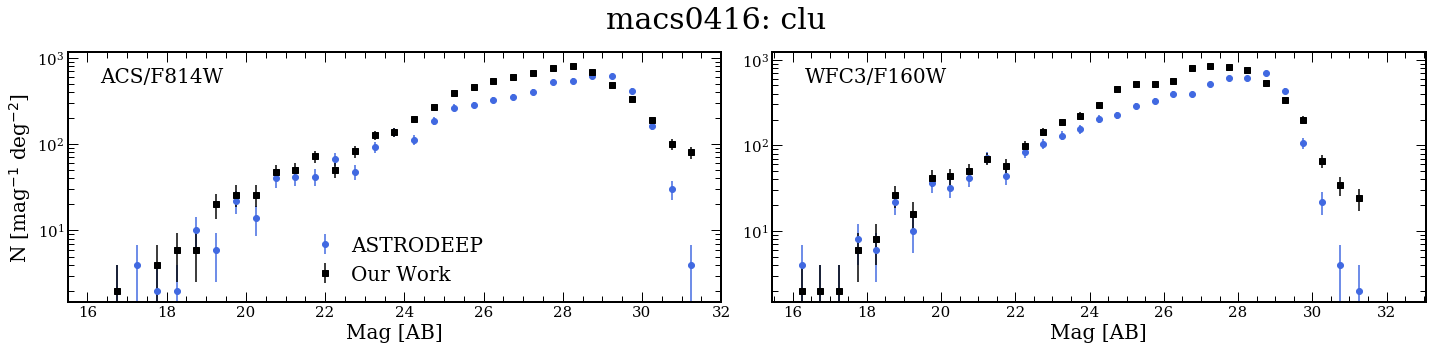}
\includegraphics[width=.95\textwidth]{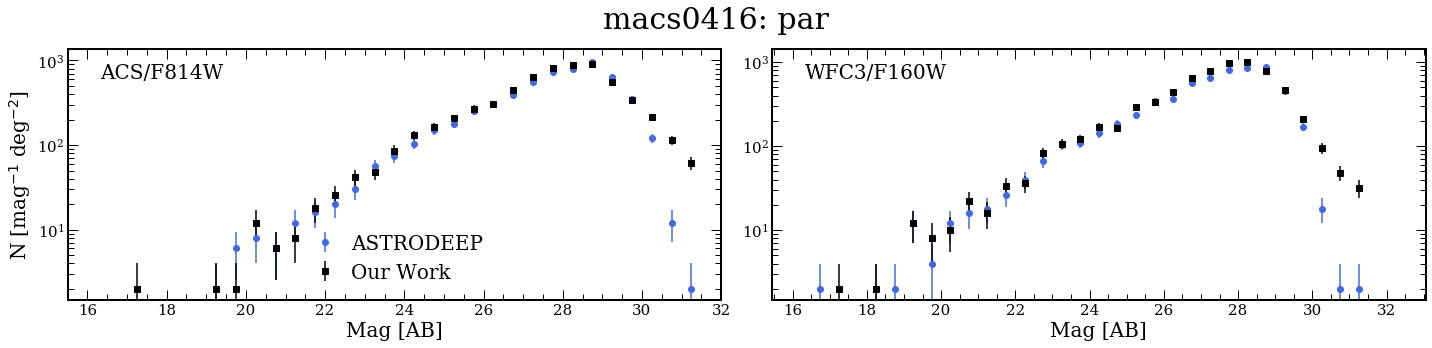}
\caption{ \textbf{Top:} The area normalized number counts in F814W and F160W filters for the central part of the Abell 370 cluster and parallel fields. We show the number counts for our catalog as black squares, while the \citet{shipley18} work is overplotted in blue circles. \textbf{Bottom:} The area normalized number counts in F814W and F160W filters for the central part of the MACSJ0416 cluster and parallel fields. We show the number counts for our catalog as black squares, while the ASTRODEEP data are overplotted in blue circles.}
\label{fig:number_counts}
\end{center}
\end{figure*}

\subsection{HFF-DeepSpace}
The HFF-DeepSpace multi-wavelength photometric catalog is presented in \citet{shipley18}. The catalog focuses on the six Frontier fields, plus the associated parallels. Our overlap with HFF-DeepSpace includes all cluster and parallel regions apart from MACS J0717.5+3745, which we will exclude from our comparison. The UV-optical photometric catalog consists of up to 17 filters with \textit{HST}/ACS and \textit{HST}/WFC3, VLT/HAWK-I $K_s$ filter, and the (post-cryogenic) IRAC 3.6 $\mu$m and 4.5 $\mu$m data, plus the archival IRAC 5.8 $\mu$m and 8 $\mu$m measurements, where available.
\par
The HFF-DeepSpace methodology starts with recombination of all the background subtracted \textit{HST} exposures from different epochs, and performing an initial image cleaning from artifacts and cosmic rays. After that, for each cluster field, they follow the \citet{ferrarese06} method to model and subtract out BCGs and the ICL from each field, performing additional background cleaning on the BCG - subtracted mosaics. This procedure allows to identify more objects magnified by the gravitational potential of the cluster members, and to assign the correct flux values to galaxies that are located close to the cluster on the sky.
\par
After the BCGs have been subtracted, all shorter wavelength mosaics are then PSF matched to the F160W band. The mosaics for all available \textit{HST} filters are then combined into a single weighted mean detection image. The source detection itself, is then performed on the resultant image by using \textsc{SExtractor}, generally following the methodology described in \citet{Skelton14}. The \textit{HST} aperture photometry extraction within a diameter of 0\farcs7 is then performed on the detection and individual PSF matched \textit{HST} images, and the correction to the total flux density is calculated from the curves of growth.
\par
Low resolution IRAC photometry is extracted by using \textsc{MOPHONGO} \citep{labbe13,labbe15}, a code developed to process longer wavelength bands, specifically focusing on potentially blended objects. Similarly to \textsc{golfir} the high resolution detection image is convolved with low-resolution kernel, and used as a model to fit the IRAC photometry. However, these models are not used to extract the flux density, but rather to correct for the possible contamination from the neighbouring sources, with the IRAC flux densities themselves being extracted from 3\farcs0 apertures. Additional flux density corrections are performed by using the PSF curves of growth. 
\par
Apart from the aperture corrections, the measured flux densities of \citet{shipley18} have undergone a number of additional modifications with the aim of providing the best possible results during SED fitting. These include the zeropoint and the Milky Way extinction 
corrections. The values presented in our photometric catalog do not incorporate such corrections, and therefore we had to de-apply them, as specified in the \citeauthor{shipley18} catalog documentation. Our knowledge of galaxy SEDs is still limited, especially as we move into the high-$z$ Universe, thus preventing us from accurately calculating the zeropoint corrections (e.g. see \citealt{brown14}), with the template error function largely alleviating these effects, however not in their entirety \citep{brammer08}. Moreover, calculating extinction corrections would largely depend on the the adopted models, introducing unwanted shifts if not done carefully. Due to this, although both the zeropoint and MW corrections are largely inconsequential ($\mathcal{O}\sim 2 \% $), we take great care to convert total flux densities back to their original values to carry out a robust and unbiased photometry comparison without these corrections.
The flux densities presented in the \citeauthor{shipley18} catalog have been additionally normalized to AB=25, rather than AB=23.9, which we also take into account when comparing the results.
\par
The total combined science area of the CHArGE/\textsc{golfir} \textit{HST} and IRAC data in our catalogue, after removing masked regions, is equal to 312.6 arcmin$^{2}$, roughly double the size of the HFF-DeepSpace coverage of 136.7 arcmin$^{2}$. This results in a higher number of objects recovered in our work, which after removing potential spurious and masked sources totals to 125,947 across all five Frontier Fields, $\times 2.3$ larger than the \texttt{use\_phot} HFF-DeepSpace sample which numbers 55,579 objects. To perform the photometric comparison between the two catalogs, we only focus on the regions where the two overlap. However, even for the parts where our data cover the same area, the CHArGE images include new exposures, mainly from the BUFFALO survey \citep{steinhardt20}, not originally present in the \citeauthor{shipley18} analysis, particularly on the edges of the old mosaic. Therefore, blind matching of all available sources would be inappropriate for a fair photometry comparison, with the variation in depth potentially introducing some deleterious effects. Instead, we focus on the central parts of both the cluster (clu) and parallel (par) regions of all five Frontier Fields, where the coverage is deepest and uniform for both catalogs. In addition, as mentioned in \autoref{sec:phot}, we do not model or subtract the BCG when computing the photometry, therefore focusing only on the central area will allow us to contrast how BCG modelling affects the final measured flux densities. We have used a 1\farcs0 matching radius, and find that the average astrometric offset between the two catalogs in equal to $\sim 0.12$ arcsec, which we have corrected for.
\par
In \autoref{fig:deltamag} we show a comparison between ALCS CHArGe/\textsc{golfir} and HFF-DeepSpace catalogs for the Abell 370 field. We carry out the comparison for individual filters, and colors separately for the cluster and parallel parts, depending on the availability of the photometry. For \textit{HST} photometry the agreement is largely good to excellent. The median offsets vary with instrument, from -0.06 mag in \textit{HST}/ACS, to 0.02 mag in \textit{HST}/WFC3 filters. Our flux densities are consistently brighter, however this does not seem to stem from the lack of BCG subtraction, but rather PSF matching performed on all bands apart from F160W, where our median difference is zero. This notion is further reinforced by the fact that the disparity between flux densities is larger for ACS filters, where the effects of PSF matching would be most noticeable. For the longer wavelength IRAC bands, we however note a much larger offset, of -0.17 mag, both for the cluster and the parallels. In an attempt to understand the discrepancy, we have first compared the science images themselves, by randomly placing 1,000 apertures in the mosaics. The extracted flux densities were different by at most $\sim 0.05$ mag, which does not explain the difference that we see in our comparison. We suspect that the discrepancy in the IRAC bands can be attributed to the difference in chosen methodology, specifically the aperture to total flux correction. 
\par
Comparison of colors yields even better results for both \textit{HST} and IRAC bands as it largely ignores any differences caused by PSF matching the individual bands, and on the aperture to total flux corrections. We show these results in \autoref{fig:delta_color}, again both for cluster and parallel parts of Abell 370. Apart from the F160W $-$ 3.6 $\mu$m color, we find offsets of the order of -0.01 mag. Despite the difference in flux measurements, the agreement between colors indicates that measurements in both catalogs have been executed in a consistent way. 
\par
Finally, in the top panel of \autoref{fig:number_counts}, we show the area weighted number of objects, as a function of their apparent magnitude. The source number counts is one of the basic tests to evaluate and characterise sensitivity-limited samples. The comparison is performed in the same "deep" areas of the Abell 370 field, for two representative ACS and WFC3 filters. We do not carry out the same analysis for the IRAC bands, as the flux offsets are too large to compare the number counts in a consistent manner. For this comparison we have only used the "clean" photometry, i.e. \texttt{use\_phot=1} flag for HFF-DeepSpace and \texttt{n\_masked=0} for our data. For each Frontier Field, our detection algorithm recovers roughly 30 \% more sources, specifically in the faint end. A small fraction of these are expected to be spurious, or are a result of overly aggressive deblending of nearby galaxies. The number counts between both catalogs are however largely consistent, and show similar depths.

\begin{figure*}
\begin{center}
\includegraphics[width=.98\textwidth]{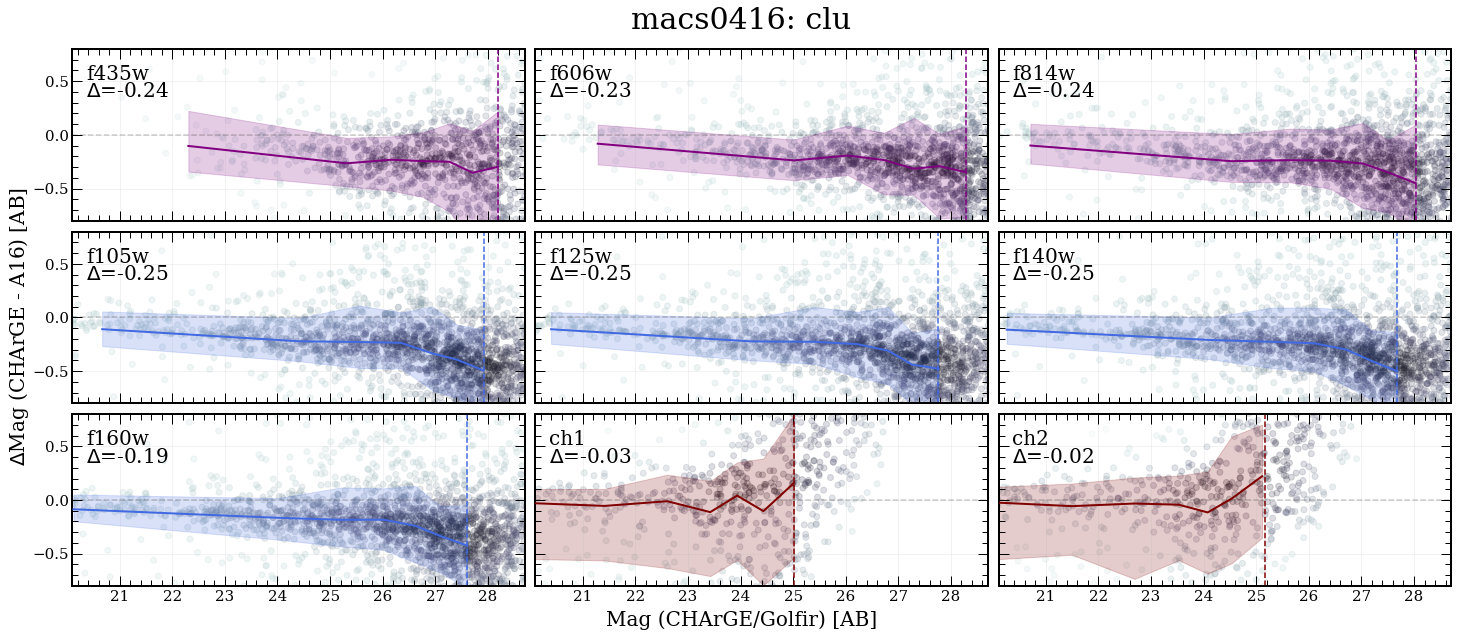}
\includegraphics[width=.98\textwidth]{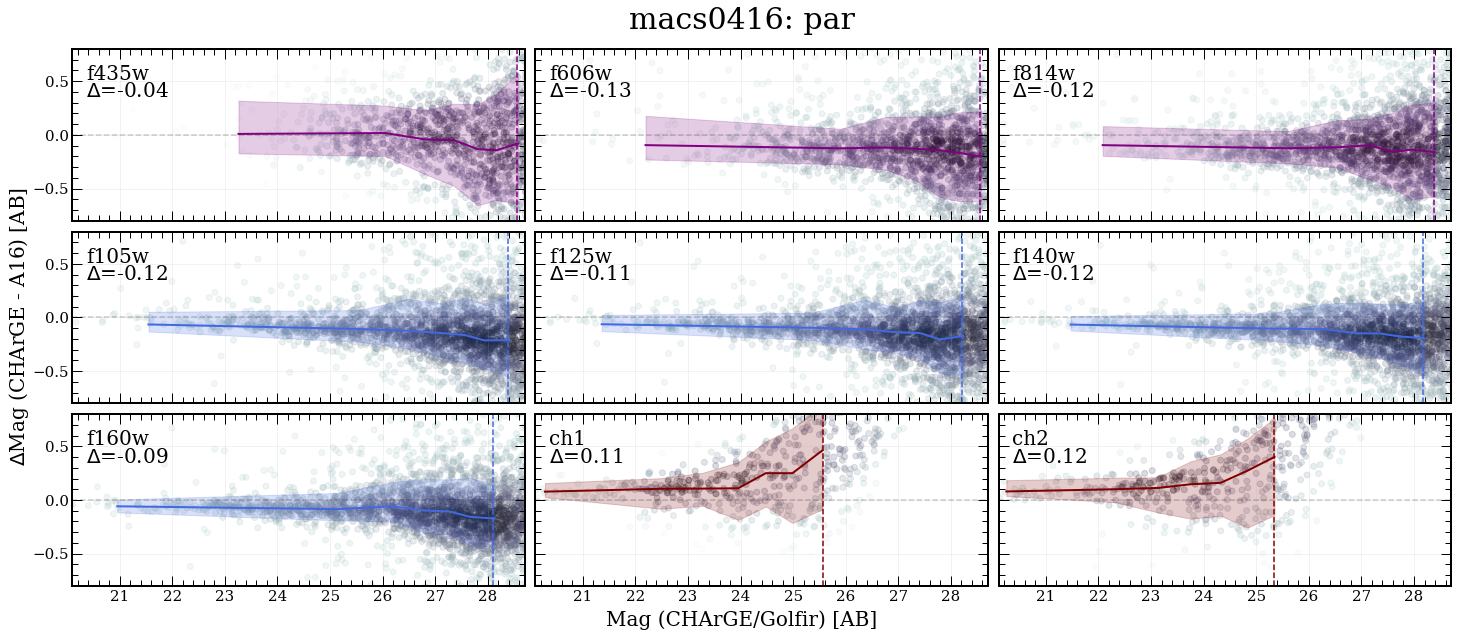}
\caption{The difference between broad-band magnitude measured in our catalog and the ASTRODEEP data,
for the cluster (top) and parallel (bottom) parts of MACSJ0416 field. Symbols and colors are the same as those in \autoref{fig:deltamag}.}
\label{fig:deltamag_astrodeep}
\end{center}
\end{figure*}

\begin{figure*}
\begin{center}
\includegraphics[width=.9\textwidth]{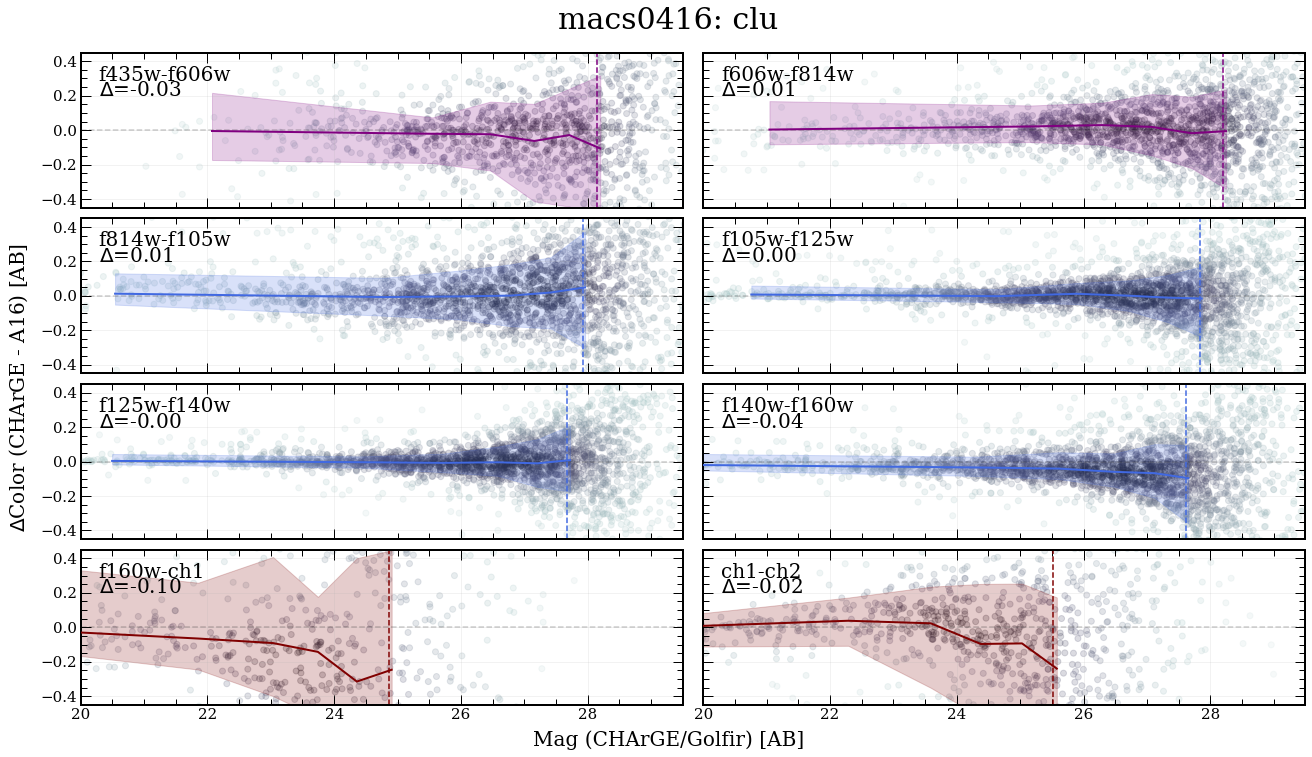}
\includegraphics[width=.9\textwidth]{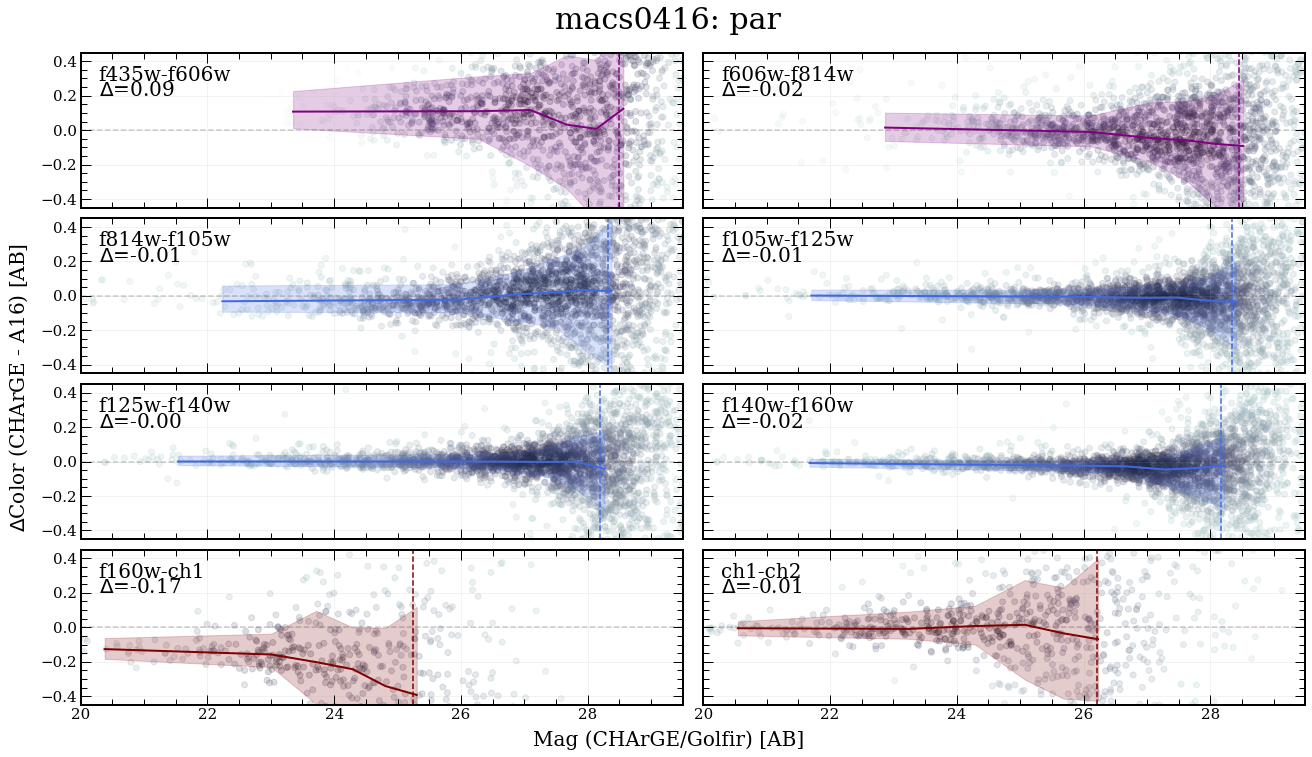}
\caption{The difference between broad-band color computed in our catalog compared to ASTRODEEP, for the cluster (top) and parallel (bottom) parts of MACSJ0416 field. The colors and symbols are the same as those in \autoref{fig:deltamag}.}
\label{fig:delta_color_astrodeep}
\end{center}
\end{figure*}

\subsection{ASTRODEEP}
For the next comparison we focus on the ASTRODEEP catalog presented in \citet{astrodeep1}, including the additional data release by \citet{astrodeep2}. The catalogs present the data for the cluster and parallel pointings for four Frontier Fields, Abell-2744, MACSJ0416, MACSJ0717 and MACSJ1149, three of which lie in the ALCS area. The photometric coverage includes 10 bands, covering \textit{HST}/ACS and WFC3, as well as HAWK-I $K_s$, and IRAC 3.6 $\mu$m and 4.5 $\mu$m data.
\par
The methodology for the ASTRODEEP catalog is largely similar to the one presented in \citet{shipley18}. The ICL and BCGs are modelled and subtracted from the $H$-band image by using models from \citet{ferrarese06}, in conjunction with \textsc{GALFIT}. The \textit{HST} source detection is performed on a single image - the WFC3 F160W band by using \textsc{SExtractor} in both \texttt{HOT+COLD} modes (e.g. see \citealt{galametz13}). The authors then use a sequential approach to subtract the ICL and BCGs from all the other cluster images, by using the output of the previous \textsc{GALFIT} run on a redder band, as an initial guess for the bluer one. All of the images are then PSF - matched to the F160W filter, by using a convolution kernel. The final \textit{HST} photometry is extracted with \textsc{SExtractor} running in dual mode, to measure the aperture and isophotal fluxes. Total fluxes in the F160W filter are computed from the 
\textsc{SExtractor} \texttt{FLUX\_AUTO} parameter. For other bands the total flux is derived from the scaling of the detection band to all the relevant isophotal colors.
\par
The low resolution IRAC photometric measurements in the ASTRODEEP catalog follow a method similar to the one outlined in this work, and \citet{shipley18}. The authors use \textsc{T-PHOT} \citep{merlin15}, which follows the same methodology as both \textsc{MOPHONGO} and \textsc{golfir}, by using high-resolution images, convolved with a low resolution PSFs to act as models for the photometric data. 
\par
As before, we only compare sources in the central parts of both cluster and parallel fields.  For that we
used a matching radius of 1\farcs0. The average astrometric offset between our sources and the ASTRODEEP catalog is $\sim$0\farcs22. In \autoref{fig:deltamag_astrodeep} we show a comparison between ALCS CHArGe/\textsc{golfir} and ASTRODEEP catalogs for the MACS0416 field. For \textit{HST} photometry the difference is quite substantial, with a median offset equal to -0.21 mag for the cluster and -0.11 mag for the parallel field, without dependence on the instrument. We suspect that the BCG subtraction is responsible for half of the offset, as the differences in flux are lessened in the parallel field. The fluxes presented in the ASTRODEEP catalog are also extinction corrected, however we believe that this is not a major contributor to the difference that we find. If we ignore the potential effects of BCG subtraction, in most cases the difference between the two catalogs is not a systematic shift, but rather an offset increasing at fainter magnitudes. The cause for the flux-dependent behavior of the offset likely originates from a different aperture used to extract the photometry, and the methodology to convert that flux to total.  Unfortunately we could not carry out the exact flux comparison to the ASTRODEEP catalog, due to the unavailability of documentation regarding the aperture sizes used, and total flux corrections. In the IRAC 3.6 and 4.5 $\mu$m filters, however, we find a remarkable agreement, of $\sim -0.01$ mag, in both cluster and parallel fields.  The color comparison, again, yields consistent results, across all \textit{HST} and IRAC filters. We find an average offset of -0.01 mag for all colors, apart from the F160W $-$ 3.6 $\mu$m color. 
\par
Despite large offsets in the ACS and WFC3 bands, we show a number count comparison to the ASTRODEEP catalog in the bottom panel of \autoref{fig:number_counts}, as the magnitude offset are too large to yield reasonable results. Notably, however, we again recover roughly 30 \% more sources, in both cluster and parallel fields on the faint end.

\subsection{CLASH \textit{HST} Catalog}
For our final broad-band photometry comparison we focus on the 
CLASH \textit{HST} catalog presented in \citet{molino17}. The catalog contains photometry for 25 massive galaxy clusters in CLASH, and overlaps with all 12 ALCS CLASH clusters that we analyzed in our work. The photometric coverage includes 14 bands from \textit{HST} only, covering \textit{HST}/ACS and WFC3/UVIS and WFC3/IR.
\par
Similarly to the HFF - DeepSpace and ASTRODEEP catalogs, the ICL and BCG contribution are subtracted from the image. This is achieved by first running \textsc{SExtractor} on the deep NIR detection images, and then based on the detected source catalog, modelling and subtraction are performed. After this all short wavelength mosaics are then PSF - homogenized to the WFC3/IR camera. The photometry itself is performed by using two different sets of apertures which the authors define as the \texttt{restricted} and the \texttt{moderate}. In the case of the \texttt{restricted} aperture \textsc{SExtractor} is forced to define the \texttt{AUTO} magnitude to the smallest available radius, which does not necessarily integrate all of the light from the galaxy, but results in the higher overall S/N. The authors argue that this approach is less sensitivity to PSF variations across images and results in more robust recovery of colors, thus yielding more accurate photo-$z$ estimates. The latter case of the \texttt{moderate} aperture is more similar to our approach with CHArGE, and relies on increased aperture sizes which aim to integrate all of the light from the galaxy. In contrast to the \texttt{restricted} apertures, this approach is argued to be more appropriate for an unbiased extraction of the physical parameters, such as stellar masses, ages and metallicities. For the purposes of this comparison we will only focus on the photometry extracted from the \texttt{moderate} apertures, as this most closely resembles the approach that we undertook in this work. 
\par
\begin{figure*}
\begin{center}
\includegraphics[width=.9\textwidth]{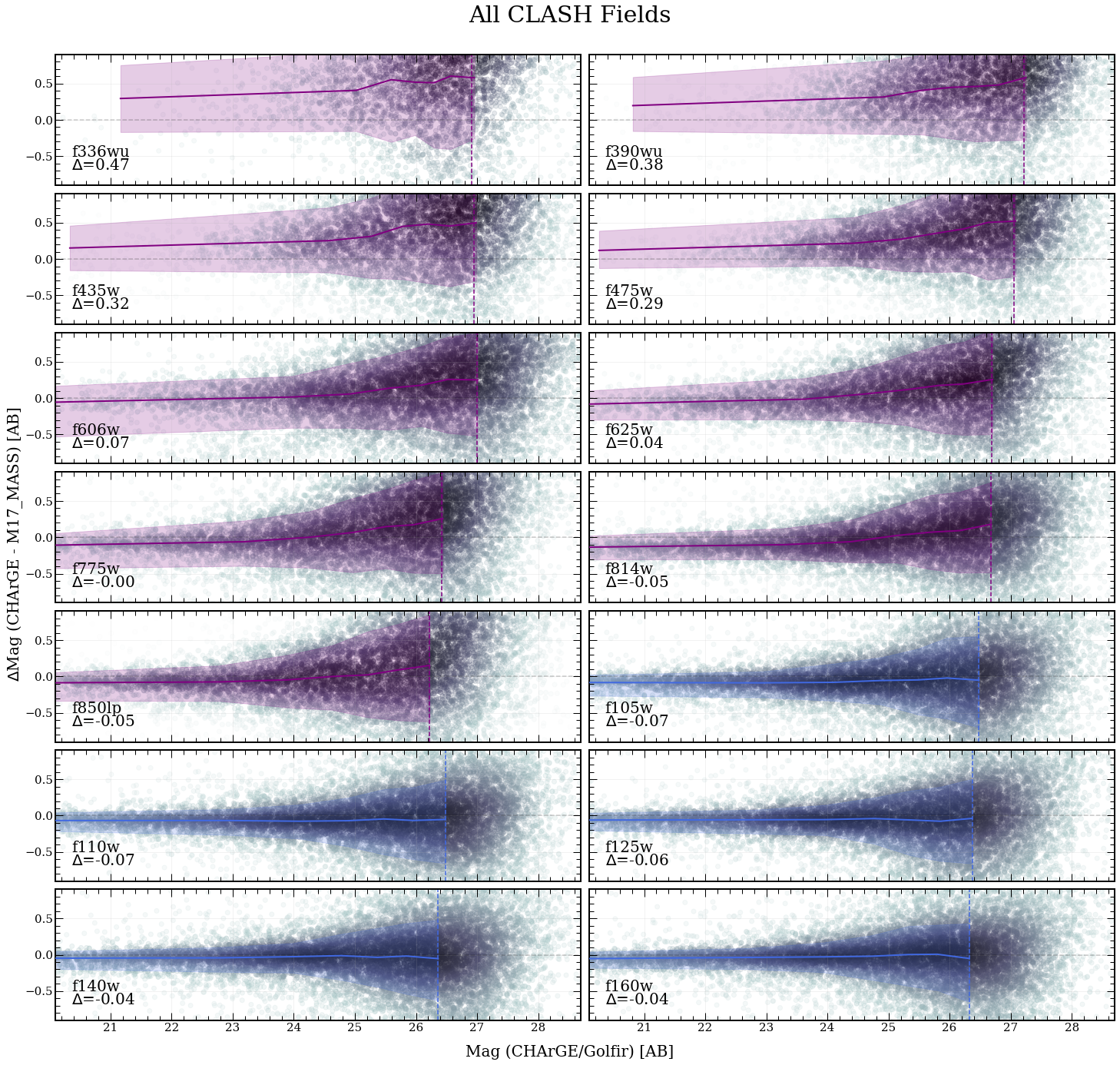}
\caption{The difference between broad-band magnitude computed in our catalog compared to CLASH photometric catalog of \citet{molino17}. The colors and symbols are the same as those in \autoref{fig:deltamag}.}
\label{fig:delta_mag_molino}
\end{center}
\end{figure*}
\begin{figure*}
\begin{center}
\includegraphics[width=.9\textwidth]{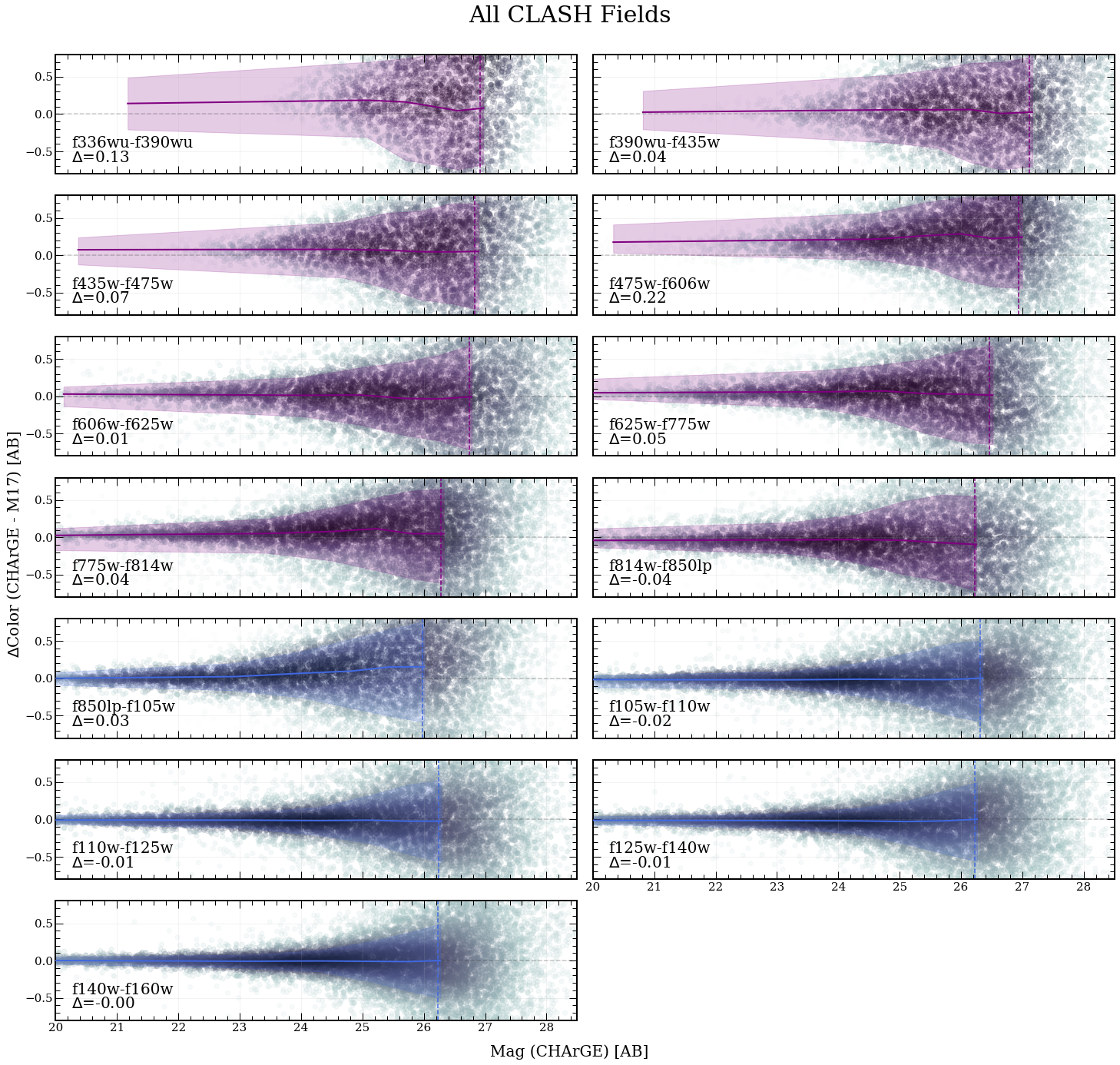}
\caption{The difference between broad-band color computed in our catalog compared to CLASH photometric catalog of \citet{molino17}. The colors and symbols are the same as those in \autoref{fig:deltamag}.}
\label{fig:delta_color_molino}
\end{center}
\end{figure*}

We match all 12 CLASH fields where CHArGE and the \citeauthor{molino17}
catalog data overlap, with a 1\farcs0 matching radius. The average astrometric offset computed across 12 fields is equal to $\sim 0.18$ arcsec. We show the comparison between the broad-band photometry for 14 \textit{HST} filters in \autoref{fig:delta_mag_molino}.
We find that the agreement between the two data-sets is mostly good, apart from the bluest \textit{HST}/ACS and UVIS bands. The median offsets are generally consistent across all instruments and vary from -0.04 mag to 0.07 mag. For all of the \textit{HST}/WFC3 and half the \textit{HST}/ACS our photometry seems to be systematically brighter, which changes towards being fainter when we move towards bluer filters.
The offsets become substantially pronounced for the bluest ACS and WFC3/UVIS bands, with medians reaching upwards of 0.47, indicating that our flux density measurements for those bands are $\sim 1.5\times$ fainter. Similarly to our comparison with the \citeauthor{shipley18} data, the fact that this discrepancy only manifests itself in a specific subset of bands indicates that its origins are not coming from the BCG or ICL subtraction. We thus conclude, as before, that the magnitude offsets are coming from the PSF matching procedure. The \texttt{moderate} aperture sizes described in \citeauthor{molino17} are optimised to integrate almost all of the light around the galaxy, where the maximum aperture size is dictated by a S/N threshold. While some considerations regarding contamination from the neighbouring have been made, the convolution of the high-resolution blue ACS and UVIS bands could have had
a deleterious effect on the extracted photometry. The color comparison, presented in \autoref{fig:delta_color_molino} yields consistent results, across the majority of \textit{HST} bands, and only strongly deviates for the aforementioned bands where the magnitude offsets are also large.

\clearpage

%





\bibliographystyle{aasjournal}
\bibliography{refs}



\end{document}